\documentclass[journal=jctcce,manuscript=article,layout=traditional]{achemso} 
\usepackage[T1]{fontenc} 
\usepackage[version=3]{mhchem} 
\usepackage[dvipsnames]{xcolor}
\usepackage{graphicx,subcaption}
\usepackage{amssymb}
\usepackage{overpic}
\usepackage{braket}
\usepackage{comment}
\usepackage{soul}
\usepackage{bm}
\usepackage{hyperref}
\hypersetup{pdfborderstyle={/S/U/W 0.5}}
\usepackage{booktabs}
\usepackage{enumitem}
\usepackage{multirow}
\usepackage{tikz}
\usepackage{svg}
\usepackage{bibunits}
\newcommand{\circled}[1]{\tikz[baseline=(C.base)]\node[draw,circle,inner sep=0.8pt](C){#1};}

\newcommand{\bX}{{\bf X}}
\newcommand{\bP}{{\bf P}}

\author{Linqing Peng}
\affiliation{Department of Chemistry, Princeton University, Princeton, New Jersey 08544, United States}
\author{Titouan Duston}
\affiliation{Department of Chemistry, Princeton University, Princeton, New Jersey 08544, United States}
\author{Nadine Bradbury}
\affiliation{Department of Chemistry, Princeton University, Princeton, New Jersey 08544, United States}
\author{Mansi Bhati}
\affiliation{Department of Chemistry, Princeton University, Princeton, New Jersey 08544, United States}
\author{Xuecheng Tao}
\affiliation{Department of Chemistry, University of Pennsylvania, Philadelphia, Pennsylvania 19104, USA}
\author{Michael Rosen}
\affiliation{Department of Chemistry, Princeton University, Princeton, New Jersey 08544, United States}
\author{Joseph E. Subotnik}
\email{subotnik@princeton.edu}
\affiliation{Department of Chemistry, Princeton University, Princeton, New Jersey 08544, United States}
\title{A Conceptual Shift In Our Understanding of Degenerate Radical Spin Systems: Spin-Rotation Coupling Turned On Its Head}

\defaultbibliography{reference}

\begin{document}
\begin{bibunit}

\begin{abstract}
    For most chemists, Kramers' degeneracy refers to the fact that for any radical system, every potential energy surface is at least doubly degenerate  (with spin up and spin down, time-reversed solutions) for all nuclear positions $\mathbf{X}$. That being said, as is well-known to the community of spin chemists, one can experimentally detect a splitting of almost every rotational energy level for a doublet system --- highlighting the fact that nuclear motion   breaks the spin degeneracy of such BO electronic states. Thus, as far as predicting experimental spectra, the implications of BO degeneracy are very limited unless one further includes a complete treatment of nuclear-electronic entanglement in a robust fashion; indeed, understanding radical molecules (and the degeneracy of their stationary states) can be extremely non-intuitive within the paradigm of  Born-Oppenheimer potential energy surfaces.  Now, as an alternative to BO theory, recent theory has suggested characterizing radical potential energy surfaces as functions of both nuclear position $\bf X$ and nuclear momentum $\bf P$, an approach which has been shown to recover a host of observables  outside of BO theory, e.g., vibrational circular dichroism, Raman optical activity, and lambda doubling.  Here, we show that such a technique predicts that {\em different spin states will follow different (nondegenerate) potential energy surfaces} and that the differences in these spin-dependent surfaces is  quantitatively consistent with experimental spin-rotation couplings -- all without any  contradiction with regard to Kramers' degeneracy. Thus, the present finding suggests there is still a great deal to learn about spin-resolved molecular reactivity, demanding a conceptual shift in our understanding of coupled spin-nuclear motion, especially in the context of chiral molecules and materials  where spin-separation is known to arise.
\end{abstract}

\section{Introduction}
For close to a century, physicists and chemists have known that the electronic spin  senses the current of nuclear rotation like a tiny magnet~\cite{hill1928quantum}. This spin-rotation coupling phenomenon has been quantitatively observed through both spin-splitting of rotational levels in microwave rotational spectroscopy of radicals~\cite{miller1976spectroscopy, davis1997jet,schuder_sub-doppler_2017} and viscosity-dependent spin relaxation in magnetic resonance\cite{atkins1966esr};  theoretical models have been developed\cite{van1951coupling, curl_jr_relationship_1965,bunker1979molecular} with {\em ab initio} applications\cite{tarczay2010first,gauss1996perturbation}. Curiously, this spin-rotation effect has consistently been excluded from  standard quantum chemistry treatments, which are always centered on the Born-Oppenheimer approximation~\cite{born1996dynamical,born1927quantentheorie} that ignores the impact of nuclear momentum on electrons and parametrizes the electronic Hamiltonian $H_\mathrm{BO}(\bX)$ and its eigenvalue potential energy surfaces $E(\bX)$ by the nuclear position.
For example, according to a BO perspective, the degeneracy of the ground state of a radical (Kramers' degeneracy) is encoded in the degeneracy of $E(\bX)$. Yet, this apparent degeneracy is clearly deceiving on some level because at low-energy/temperature scale (for realistic molecules with moving nuclei),
the aforementioned
spin-rotation coupling correlates electronic spin with nuclear rotation and splits the doublets into nondegenerate electronic-nuclear-coupled states (thus apparently lifting degeneracy). Clearly, the BO perspective is limiting, and for a more thorough discussion of Kramers' degeneracy, please see Sec.~\ref{sec:conclusion} below. Moreover,
a complete understanding of the degeneracy and degeneracy lifting of quantum states involving radicals is essential for modern technology where  precise control of low-energy spin excitations is utilized in an increasingly wide range of fields, spanning across quantum computing and sensing~\cite{gaita2019molecular,atzori2019second, bayliss2020optically}, energy-efficient spintronics~\cite{bogani2008molecular, thiele2014electrically,christou2000single, sessoli2017magnetic}, and molecular chirality~\cite{bloom2024chiral,ray1999asymmetric,adhikari2023interplay}. In some cases, spin-nuclear coupling has even been the bottleneck to prolonging spin coherence in quantum spin devices~\cite{goodwin2017molecular,zhang2025spin,mirzoyan2020dynamic}. For all of these reasons, there is a strong reason for us to shift our view of radical chemistry away from the very limiting BO perspective.

Is there a simple means to go beyond the limits of the BO approximation and characterize molecules and materials in such a fashion as to capture the nuclear rotational and vibrational effects on electronic spins?  Or is the BO framework (which itself is almost a century old\cite{born1927quantentheorie}) the only way forward? 
As noted above, the main approximation of BO theory is that we parameterize molecular orbitals and molecular energy parametrized by nuclear position $\bX$.
Over the last few years, we have come to realize that
a more accurate description of electrons and spin can be achieved by parameterizing the electronic Hamiltonian not only on nuclear position $\mathbf{X}$ but also on nuclear momentum $\mathbf{P}$~\cite{bian2025review,tao2025basis,qiu2024simple,bradbury2025symmetry}, a so-called ``phase space electronic structure'' Hamiltonian, $H_\mathrm{PS}$.  Working with $H_\mathrm{PS}$, rather than $H_\mathrm{BO}$, one is guaranteed to conserve linear and angular momentum, and from this construction, the physics of electron-nuclear rotation coupling, i.e., exchange of angular momenta between electrons and nuclei, also naturally arises. To date, we have shown that for chiral molecules, this method can predict vibrational circular dichroism~\cite{duston2024phase,tao2024electronic} and Raman optical activity~\cite{tao2025non}; for diatomic molecules, this method near quantitatively predicts $\Lambda$-doubling~\cite{peng2026phasespaceelectronicstructure}; for model systems, this method can even improve predictions of vibrational energies~\cite{bian2025phase,wu2025recovering}. In this work, we will now show that the same phase space approach is capable of predicting the electronic spin-rotation coupling that is general to all open-shell molecules. In other words, working with 
$H_\mathrm{PS}(\mathbf{X},\mathbf{P})$ rather than $H_\mathrm{BO}(\bX)$, one finds a $6N$ (not $3N$) dimensional potential energy surface $E(\bX,\bP)$  that incorporates all of  BO theory
while also
quantitatively capturing spin-rotation coupling and a host of other effects that break BO theory.  The result is an intuitive framework to visualize and understand the nuclear effect on electrons,  with roughly the cost of a standard calculation (provided one includes spin-orbit coupling).  Most importantly, the approach can scale very well with system size. Thus, below we focus on fundamental chemistry for small molecules, but there is every reason to think that, going forward, this same method will shed light on phenomena in large systems such as Einstein-de Haas physics, the Barnett effect, chiral-induced spin selectivity, and superconductivity. As such, we argue that, when looking to understand the chemistry of radical spin systems and/or other systems with degeneracies or near degeneracies, the time has come for a conceptual shift in our approach to electronic structure.

This article is outlined as follows. First, we will review the spin-rotation coupling following the standard perturbation and model Hamiltonian treatment in the literature. Second, we will briefly review phase space electronic structure theory and the physical meaning of the phase-space potential energy surface (PES). Thereafter, we will show formally how phase space electronic structure theory, combined with a 3D rotational model, is capable of capturing the same coupling without the need to sum over all excited states. To demonstrate the power for experimentalists, we present numerical benchmarks of the phase space theory on the prediction of spin-rotation coupling of four molecules and demonstrate that phase space theory predictions are quantitatively accurate. Finally, we conclude by showing how and why phase space electronic structure theory necessarily gives us remarkable new insight into the meaning of degeneracy (e.g., Kramers' degeneracy) for coupled nuclear-electronic systems, and we point out many future directions.

\section{The Standard Born-Oppenheimer View of Spin-Rotation Coupling}
\label{sec:perturbation}
Let $L$ be the electronic angular momentum, and let $R$ be the rotational/orbital angular momentum of the nuclei.
Electronic spin-nuclear rotation coupling is the effect that splits degenerate states with the same spin angular momentum $S$ and total rotational angular momentum $N$ ($\hat{\mathbf{N}} \equiv \hat{\mathbf{R}} + \hat{\mathbf{L}}$)  into nondegenerate states with different total angular momentum $J$ ($\hat{\mathbf{J}}  \equiv \hat{\mathbf{R}} + \hat{\mathbf{L}} + \hat{\mathbf{S}}$). Such a splitting originates from two interactions: 1) a first-order relativistic spin-nuclear-orbit coupling $E^{\mathrm{SR(1)}}$ whereby the electronic spin feels the magnetic field of a nearby rotating nucleus 
\begin{equation} \label{eq:SR1}
    H^\mathrm{SR,(1)} = -\frac{a_0g\alpha^2}{2m_e}\sum_{Ai} \frac{Z_A}{|\hat{\mathbf{r}}_i - \mathbf{X}_A|^3}(\hat{\mathbf{r}}_i - \mathbf{X}_A) \times \frac{\mathbf{P}_A}{M_A} \cdot \mathbf{\hat{S}}_i
\end{equation}
where $a_0$ is the Bohr radius, $g$ is the Land\'e $g$-factor, and $\alpha$ is the fine structure constant,
which is a type of spin-other-orbit interaction, and 2) a more complicated second-order coupling with energy $E^{\mathrm{SR(2)}}$ due to the double perturbation of the electronic SOC $ \hat{H}_\mathrm{SOC}$  and the electronic orbital Coriolis potential $-\mathbf{\omega} \cdot \hat{\mathbf{L}} = -\mathbf{I}^{-1} \mathbf{N} \cdot\hat{\mathbf{L}}$  in the rotating nuclear frame. 
Here,  $\mathbf{\omega}$ represents the angular velocity of the molecular frame and $I$ is the nuclear moment of inertia.
For all but a few very small systems, e.g., H$_2$, 
the first effect is small compared to the second effect and thus, theoretical discussions here focus on the second effect. The (larger) energy splitting due to the second effect can be modeled with second-order perturbation theory. 

The calculation of spin-rotation coupling proceeds by calculating all (low-energy) unperturbed BO electronic eigenstates $|i\rangle$, where $|0\rangle$ is the ground state.  A double perturbation then  yields
the following second-order energy correction: 
\begin{equation}
    E^{\mathrm{SR} (2)}  = \sum_{i\neq 0} \frac{2\langle 0|\hat{H}_\mathrm{SOC}|i\rangle \langle i|-\mathbf{I}^{-1} \mathbf{N} \cdot\hat{\mathbf{L}}|0\rangle}{E_{0} - E_i} = - \mathbf{I}^{-1}\mathbf{N}\cdot 2\langle \Psi^{(1)}_\mathrm{SOC}|\hat{\mathbf{L}}|0\rangle \approx -\mathbf{I}^{-1}\mathbf{N}\cdot \langle \Psi^{(1)}_\mathrm{SOC}|\hat{\mathbf{L}}|\Psi^{(1)}_\mathrm{SOC}\rangle \label{Eq:E2_PT}
\end{equation} 
Here $|\Psi^{(1)}_\mathrm{SOC}\rangle$ represents the first-order corrected wavefunction of state $|0\rangle$ under the SOC perturbation, and the second equal sign assumes $\mathbf{L}=0$ for nonrelativistic BO ground states  (i.e. $\left< 0 \middle| L_\mu \middle | 0 \right> = 0$ for all $\mu=a,b,c$), which is true for molecules whose spatial part is nondegenerate.
It is crucial to emphasize that, for a doublet or triplet or any spin multiplet state, the ground state $\ket{0}$ is spin-degenerate; in particular, for a  Kramers doublet, one generates pairs of BO electronic ground states by applying the time reversal operator (which has the effect of generating opposite spin orientations).
Thus, the energy in Eq.~\ref{Eq:E2_PT} depends on the choice of $\ket{0}$.
For a given choice of $\ket{0_s}$ (which has a unique spin direction $s$), we  should really write $E^{\mathrm{SR}(2)}_s$; for this state, and the second-order energy splitting is $\Delta E^\mathrm{SR} = 2E^{\mathrm{SR} (2)}_s$. 

Eq.~\ref{Eq:E2_PT} offers a conceptually straightforward, but practically difficult or impossible, means to evaluate the energy splitting as caused by an interaction of the electronic spin and molecular rotation in the $\mathbf{N}$ direction.  The difficulty arises from the (infinite) sum over excited states. In practice, for a doublet, if one seeks a compatible model Hamiltonian of the form, 
\begin{equation}\label{eq:SR_model}
    H_\mathrm{SR} = \hat{\mathbf{S}} \cdot \mathbf{\epsilon} \cdot \hat{\mathbf{N}}
\end{equation}
with a $3 \times 3$ spin-rotational coupling tensor $\epsilon$,
the most straightforward means to evaluate this tensor is to use 
second-order response theory. More specifically, for a Hamiltonian of the form $\hat{H} = \hat{H}_0 + \lambda \hat{H}_\mathrm{SOC} - \mathbf{I}^{-1}\mathbf{N}\cdot \mathbf{\hat{L}} $, we evaluate\cite{tarczay2010first,cfour}:
\begin{equation} 
    \epsilon_{ab} = - 4 \mathbf{I}^{-1}
   \left< \frac {\partial \Psi^{(1)}_{\mathrm{SOC},a}}{\partial \lambda} \middle | \hat{L}_b \middle | 0_a\right> =  2\frac{\partial^2 E^\mathrm{SR(2)}_a}{\partial \lambda \partial N_b}.\label{eq:CC_SR}
\end{equation}
If one were to include the first-order relativistic coupling, one would simply replace $E^\mathrm{SR(2)}$ with $E^\mathrm{SR(1)} + E^\mathrm{SR(2)}$ in Eq.~\ref{eq:CC_SR}.
Alternatively, one can also approximate this coupling constant from the electronic $g$-tensor using Curl's approximate formula $\mathbf{\epsilon} = -\mathbf{I}^{-1}(g_e \mathcal{I}_3 - \mathbf{g})\hbar^2 $ where $g_e = 2.002319$ and $\mathcal{I}_3$ denotes the $3\times3$ identity matrix~\cite{curl_jr_relationship_1965,grein2004trends}. Note that  Curl's formula 
is only valid up to the second-order energy perturbation~\cite{tarczay2010first}. For an in-depth discussion of this tensor and different approaches to its calculation within BO theory, see the  Supplementary Information (SI).  
Note that, no matter how the calculation is done, within BO theory, the calculation of $\epsilon$ requires a response calculation (e.g. at the very least, one must run a coupled-perturbed HF calculation).

So far, we have calculated the spin-rotation coupling energy for a classical rotational angular momentum $\mathbf{N}$ fixed relative to the body frame, i.e., a rotation around a molecular axis parallel to the orientation of $\mathbf{N}$. This calculation allowed us to calculate the spin-rotation coupling tensor $\epsilon_{ab}$. However, in order to compare against experiment, it is crucial to emphasize that (i) only energy splittings are measurable (not $\epsilon$) and (ii)
not only do molecules rotate around a body axis, but also the body axis rotates in the space frame (i.e. the whole molecule tumbles in space end-over-end), and this entire process is quantized.   Thus, in order to compare against experiment, we require a {\em quantum} spin-rotation coupling splitting. To that end,
note that a rotational wavefunction is described by three quantum numbers: $N$, $K$ (projection on the body $c$-axis), and $M$ (projection on the space $z$-axis).
To derive the energy 
associated with a quantum mechanical 3D rotation of free molecules in space, we must diagonalize the quantum Eq.~\ref{eq:SR_model} above (where now $N$ is a quantum rotation operator); the difference in eigenvalues yields the spin-rotation energy splitting.

In the case of a symmetric top molecule, the spin-rotating tensor is diagonal in the principal axis basis, similar to the nuclear spin-rotation tensor\cite{flygare1974magnetic}. For a doublet molecule, the energy splitting between the $J=N+1/2$ state and the $J=N-1/2$ state for a given $(N, K)$ rotational state is independent of $M$ and equal to\cite{davis1997jet}
\begin{equation} \label{eq:Esr_NK}
     \Delta E_\mathrm{SR} (N, K) = -(N+1/2)\left[ \epsilon_{bb} - (\epsilon_{bb} - \epsilon_{cc}) K^2/N(N+1)\right]
\end{equation}
where $c$ is the symmetry axis and $I_a=I_b$ (see Sec.~\ref{Sec:supp_3Dmodel} in SI for derivation). This calculated energy splitting allows us to directly compare with experimental rotational spectra associated with transitions between states with different $(N,K)$ rotational levels and spins.

\section{A Phase Space Electronic Structure View of Spin-Rotation Coupling} 

The (dominant) mechanism underlying spin-rotation coupling arises from dynamics in a non-inertial frame:  nuclear rotation induces an electronic current in the rotating frame, which then creates an internal magnetic field. Consider a system with an unpaired electron. Under the relativistic BO approximation, the ground electronic states are a degenerate Kramers doublet, and suppose the unpaired electron lies primarily along the $p_c$ orbital. Now when the molecule rotates around an axis perpendicular to the $c$-axis, for example, counterclockwise around the $a$-axis, the electronic wavefunction cannot instantaneously relax to the ground BO state of the new nuclear configuration and rotates relative to the nuclear frame in the opposite direction, i.e., clockwise around the $a$-axis. Therefore, when we describe the electronic wavefunction in the rotating nuclear frame, the unpaired electron not only occupies the $p_c$ orbital, but also $p_b$ orbital partially, and generates an electric current with nonzero electronic orbital angular momentum $\mathbf{L}$ that increases linearly with the rotation, and interacts with the spin and leads to a spin-rotating splitting.

How should we include such non-inertial forces within the electronic structure? How do we solve the electronic structure problem in the frame of moving nuclei, and thus incorporate all non-inertial forces on the electrons?
The ansatz of phase space (PS) electronic structure theory is that we can take inspiration from the adiabatic representation of quantum mechanics. Mathematically, one can find (in just about any textbook\cite{born1996dynamical,takatsuka2014chemical,baer2006beyond} ) the fact that, after diagonalizing the electronic Hamiltonian, the Born-Huang framework leads to a Hamiltonian of the form:
\begin{equation} \label{eq:BH}
    [\hat{H}_\mathrm{BH}]_{ij} = \sum_{A,k} \frac{1}{2M_A}(\hat{\mathbf{P}}_A \delta_{ik} - i\hbar \mathbf{d}^A_{ik}) \cdot (\hat{\mathbf{P}}_A\delta_{jk} - i\hbar \mathbf{d}^A_{jk})  + E_i \delta_{ij}
\end{equation}
where $i,j,k$ label electronic adiabatic states and $A$ is the atom index. The $\mathbf{d}^A_{ik}$ tensor in Eq.~\ref{eq:BH} is the famous derivative coupling operator that hides all of the errors of Born-Oppenheimer theory; within BO theory, one sets $\mathbf{d}^A_{ik} = 0$. Some of the errors that arise from ignoring $\mathbf{d}$ are well known.  For instance, it is well known that the presence of off-diagonal $\mathbf{d}_{ij}$ elements leads to transitions between adiabatic surfaces $i$ and $j$.  Slightly less well known is that, for motion in one dimension 
in the adiabatic representation, the diagonal elements $\mathbf{d}_{jj}$ captures the momentum of the electrons, so that (i) $\langle\hat{\mathbf{P}}\rangle$ must be interpreted as the total nuclear plus electronic momentum and (ii) $\langle\hat{\mathbf{P}}-\mathbf{d}_{jj}\rangle$ must be interpreted as the nuclear momentum when moving along surface $j$. For more details, see Ref.~\cite{littlejohn2024diagonalizing} and Sec.~\ref{sec:supp_P} in the SI. Thus, in order to recover the relevant non-inertial forces described above, one would like to include {\bf d} in a stable fashion when solving the electronic structure problem. 

Alas, there are many difficulties in working with $\mathbf{d}$ above, none less important than the obvious fact that, in order to include $\mathbf{d}$ in an electronic structure problem, one must first solve an electronic structure problem. The resulting self-consistent problem leads to insurmountable difficulties -- both conceptually and practically. With such difficulties in mind, the ansatz of phase space electronic structure theory is to approximate the $\mathbf{d}$ tensor by a well-defined one-electron operator (properly chosen) and solve the electronic structure problem $\hat{H}_\mathrm{PS} |\psi\rangle = E |\psi\rangle$ where
\begin{align} \label{Eq:Hps}
    \hat{H}_\mathrm{PS}(\mathbf{X}, \mathbf{P}) =& \sum_{A}\frac{1}{2M_A} (\mathbf{P}_A - i\hbar \hat{\mathbf{\Gamma}}_A (\mathbf{X})) \cdot (\mathbf{P}_A  - i\hbar \hat{\mathbf{\Gamma}}_A (\mathbf{X})) + \hat{H}_\mathrm{el}(\mathbf{X})
\end{align}
In Eq.~\ref{Eq:Hps}, both $\mathbf{X}$ and $\mathbf{P}$ are c-numbers, i.e. simple parameters; they are {\em not} operators. 
Eq.~\ref{Eq:Hps} approximates the derivative coupling $\mathbf{d}_{IJ}^A$ with a one-body electronic operator $\hat{\mathbf{\Gamma}}_A (\mathbf{X})$ that preserves the key symmetry properties of $\mathbf{d}_{IJ}^A$. When only the spatial part of electrons is described in the molecular frame, this operator $\hat{\mathbf{\Gamma}}_A (\mathbf{X}) = \hat{\mathbf{\Gamma}}^\mathrm{spatial}_A (\mathbf{X})$ ensures the conservations of total electron and nuclear linear momenta and of total orbital angular momenta.
(In the language of Ref.~\citenum{bian2025review}, $\hat{\mathbf{\Gamma}}^\mathrm{spatial}= \hat{\mathbf{\Gamma}}' + \hat{\mathbf{\Gamma}}''$.)
If one also describes the electronic spin in the molecular frame, the augmented operator $\hat{\mathbf{\Gamma}}_A (\mathbf{X}) = \hat{\mathbf{\Gamma}}^\mathrm{spatial}_A (\mathbf{X}) + \hat{\mathbf{\Gamma}}^\mathrm{spin}_A (\mathbf{X}) $ will ensure conservation of total angular momentum including the electronic spin.
(In the language of Ref.~\citenum{bian2025review}, $\hat{\mathbf{\Gamma}}^\mathrm{spin}= \hat{\mathbf{\Gamma}}'''$.)
The exact form of $\hat{\mathbf{\Gamma}}$ and other symmetries it preserves can be found in Sec.~\ref{sec:supp_gamma} of the Supplementary Information.

\subsection{Interpretation of $\mathbf{P}$ and $\hat{\mathbf{\Gamma}}$ Within a Phase Space Calculation} 

One can diagonalize $\hat{H}_{PS}(X,P)$  using Hartree Fock or DFT (or any other wavefunction method) just like for standard BO theory; the only difference is that one must allow for complex-valued molecular orbitals. The more interesting component of PS electronic structure theory is the interpretation of the result.  In particular, in three-dimensional space, the meaning of $\mathbf{P}$ and its associated angular momentum $\mathbf{L}^\mathrm{n} = \mathbf{X} \times \mathbf{P}$ depends on our choice of $\hat{\mathbf{\Gamma}}$. For instance, if we choose $\hat{\mathbf{\Gamma}}=\hat{\mathbf{\Gamma}}^\mathrm{spatial}$, such a choice would correspond to orienting the electronic orbital frame relative to the nuclei but leaving the electronic spin in the lab frame. Thus, $\mathbf{L}^\mathrm{n} = \mathbf{X} \times \mathbf{P}$ would represent the nuclear plus electronic orbital angular momentum $\mathbf{N}$. In such a case, one can interpret $i\hbar \sum_A \mathbf{X}_A \times \hat{\mathbf{\Gamma}}_A$ as electronic $\hat{\mathbf{L}}$. Vice versa, if we choose $\hat{\mathbf{\Gamma}}=\hat{\mathbf{\Gamma}}^\mathrm{spatial}+ \hat{\mathbf{\Gamma}}^\mathrm{spin}$, such a choice would correspond to orienting both the electronic orbital frame and the elecronic spin frame relative to the nuclei.  Thus, $\mathbf{L}^\mathrm{n}$ would represent the total (nuclear plus electronic orbital plus electronic spin) angular momentum $\mathbf{J}$.  In such a case, one can interpret $i\hbar \sum_A \mathbf{X}_A \times \hat{\mathbf{\Gamma}}_A$ as $\hat{\mathbf{L}} + \hat{\mathbf{S}}$. Throughout this article, we will reserve $\mathbf{L}^\mathrm{n}$ exclusively for the angular momentum associated with the phase space nuclear momentum parameter  $\mathbf{X} \times \mathbf{P}$, and use $\mathbf{R}$ to represent the physical nuclear kinetic angular momentum. In the following, we will assume the electronic spin is described in the lab frame, and correspondingly use  $\hat{\mathbf{\Gamma}}=\hat{\mathbf{\Gamma}}^\mathrm{spatial}$ unless otherwise noted.

\subsection{The Phase Space Potential Energy Surface}\label{sec:PES}

\begin{figure}[htb!]
    \centering
    \includegraphics[width=\columnwidth]{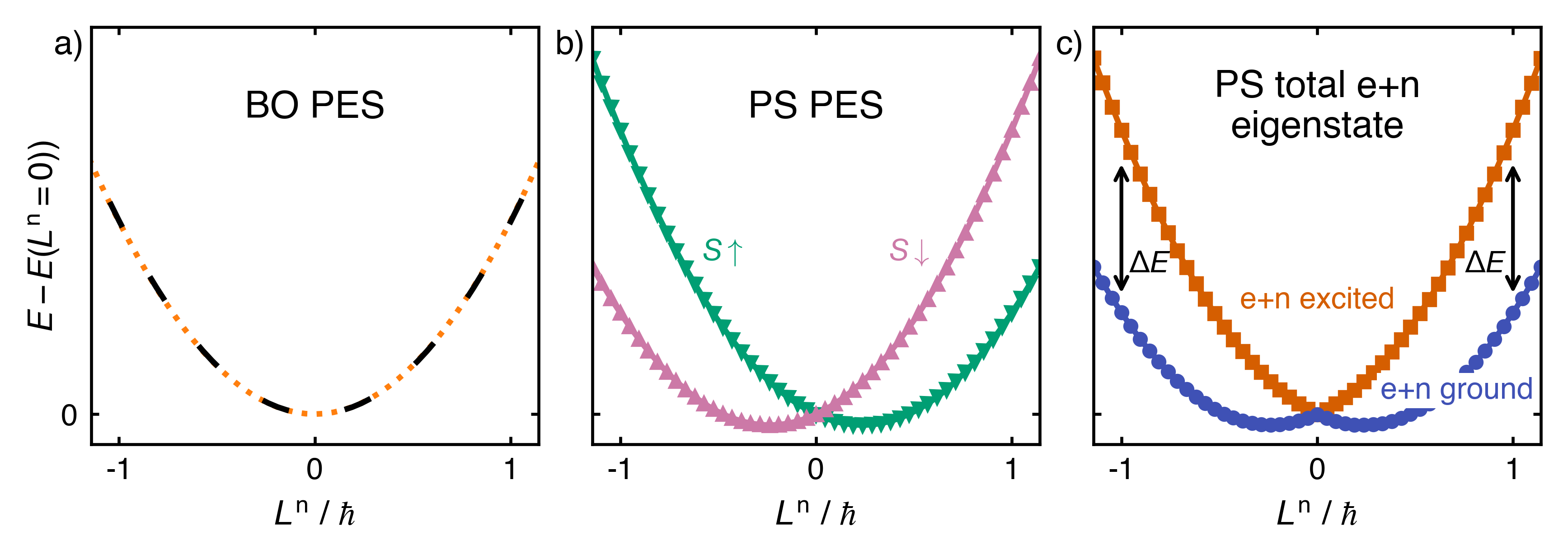}
    \caption{Lowest two potential energy surfaces under the a) BO approximation; b) the phase space theory, plotted as a function of the canonical nuclear momentum $L^\mathrm{n}$ and colored based on the spin character; c) the same phase space PES, but colored based on their relationship to the total electronic and nuclear eigenstates. Figure not to scale. Intuitively, the spin-rotation splittings arise from the difference in energy between the broken symmetry ground and excited state surfaces labeled as $\Delta E$ in c).}
    \label{fig:PES}
\end{figure}

If one sets $\mathbf{P}=0$, the PS PES along $\mathbf{X}$  resembles the BO PES extremely closely -- the $\hat{\mathbf{\Gamma}}$ term is small. At the same time, however, unlike the case for BO theory, one can find interesting features in the PS PES if one explores how the PES depends on $\mathbf{P}$.
More precisely, whereas  BO theory does not admit a coupling between nuclear momenta and the electronic degrees of freedom, leading to a PES that is separable between $\mathbf{X}$ and $\mathbf{P}$ (and also harmonic in $\mathbf{P}$), PS theory admits a much richer physics.  

As the simplest non-trivial example, consider a system with an odd number of electrons and 1/2 spin. In Figure~\ref{fig:PES}, we plot its PESs, on the left the BO PES which includes the BO kinetic energy $\sum_A \mathbf{P}_A^2/2M_A$, and on the right the PS PES which includes the PS kinetic energy $\sum_A (\mathbf{P}_A - i\hbar \hat{\mathbf{\Gamma}}_A (\mathbf{X}))^2/2M_A$ through the electronic Hamiltonian Eq.~\ref{Eq:Hps}. The BO electronic ground states are a degenerate Kramers doublet, giving rise to two degenerate BO PESs in Fig.~\ref{fig:PES}a. In contrast, the two PS PESs with opposite spin orientations in Fig.~\ref{fig:PES}b are shifted in opposite directions because they have nearly opposite $i\hbar\langle \hat{\mathbf{\Gamma}}_A\rangle$. As a result, the two PS PESs form a small energy splitting between the ground and excited states (Fig.~\ref{fig:PES}c). 
In most molecules, $i\hbar\langle \hat{\mathbf{\Gamma}}_A\rangle$ is small, so on a coarser energy scale, the PS PESs would appear nearly identical to the two degenerate BO PESs in Figure~\ref{fig:PES}a. At a fine energy scale or lower temperature, however, such spin splitting becomes significant.

\subsection{$\Lambda-$splitting and Spin-Rotation Coupling}
Ref.~\citenum{peng2026phasespaceelectronicstructure} shows that one can use the PS Hamiltonian in Eq.~\ref{Eq:Hps} and the resulting PESs in Fig.~\ref{fig:PES}c to match the $\Lambda$-splitting that occurs in the diatomic NO molecule.  For the case of $\Lambda$-splitting with a diatomic, if one ignores spin, the overall BO degeneracy arises from the degeneracy of two electronic states with different {\em electronic orbital} angular momentum (e.g. usually two states with partially filled $p_x$ and $p_y$ orbitals); this degeneracy is broken as the molecule rotates in the $xz$ or $yz$ planes. (For a discussion involving a proper treatment of spin, see Ref.~\citenum{peng2026phasespaceelectronicstructure}). 
For our purposes here, we note that the physics of broken symmetry in Fig.~\ref{fig:PES} applies equally well to spin-rotation coupling as to $\Lambda$-splitting. The only difference is that, for the former, the degeneracy arises from the electronic spin (rather than orbital degeneracy for the latter).

\subsection{Phase Space View of Spin-Rotation Coupling} \label{sec:ps_SR} 
As shown in Fig.~\ref{fig:PES}, diagonalizing a PS Hamiltonian yields two PESs that are of approximately parabolic shape with shifted minima for the two spin states of a Kramers doublet, with a mirror symmetry with respect to $L^\mathrm{n}=0$.  To recover the $\epsilon$ tensor (and the spin-rotation splitting), we assert that one can 
fit each element of the spin-rotational coupling tensor $\epsilon_{\mu \nu},  \mu,\nu=a,b,c$ 
to the PS energy splitting at the same $L^\mathrm{n}_\nu$ between two states with opposite $S_\mu$ when nuclei rotate around the $\nu$ axis: $\Delta E_{\mu\nu} (L^\mathrm{n}_\nu =N_\nu) = 2|S_\mu| \epsilon_{\mu\nu} N_\nu $. As an example, for a system with $S\approx 1/2$ and PS calculations at $L^\mathrm{n}_\nu=1$, the energy difference between the two solutions with $S_\mu \approx \pm 1/2$ should be equal to $\epsilon_{\mu\nu}$. 
Note that, according to this formalism, one does not need to invoke any response calculations (as in Eq.~\ref{Eq:E2_PT} for BO above).
An alternative (but mutually consistent) PS approach to calculating $\epsilon$ is discussed in Sec.~\ref{sec:supp_alternative_dE} of the SI, which connects more directly to the BO approach in Sec.~\ref{sec:perturbation}.  

Before concluding, there is one important nuance worth mentioning.
So far, we have discussed the PS PESs in the context of quantizing spin in the space frame, and as a result, $L^\mathrm{n}$ represents the total orbital angular momentum $N=R+L$. What if we quantize the spin in the molecular frame and use $\hat{\mathbf{\Gamma}}_A (\mathbf{X}) = \hat{\mathbf{\Gamma}}^\mathrm{spatial}_A (\mathbf{X}) + \hat{\mathbf{\Gamma}}^\mathrm{spin}_A (\mathbf{X})$? In this case, $L^\mathrm{n}$ represents the total angular momentum $J=N + S$, and one can extract the spin-rotation splitting by evaluating the energy difference between one calculated PES at $L^\mathrm{n}=N+1/2$ and  a second PES at $L^\mathrm{n}=N-1/2$.    As shown in Fig.~\ref{fig:supp_G3}, the relative error compared to values predicted without $\hat{\mathbf{\Gamma}}^\mathrm{spin}_A (\mathbf{X})$  decreases for smaller splittings. 
This trend indicates that the current PS implementation with $\hat{\mathbf{\Gamma}}^\mathrm{spin}_A (\mathbf{X})$ included yields increasingly more accurate spin-rotation couplings as the molecule becomes heavier, so that (i) there is stronger SOC and (ii) the rotational periods become longer.  This conclusion makes intuitive sense since stronger SOC must couple spin to the molecular frame more strongly and longer rotational periods allow the spin to relax to a stationary point in the spirit of a BO calculation (see also Ref. \citenum{bradbury2025symmetry}).

In the following, we choose to quantize the spin in the space frame. Consistent with this choice, $L^\mathrm{n}$ will represent $N$, and the spin-rotation coupling tensor will be fitted from the energy splitting of the lowest two PS PESs at the same $L^\mathrm{n}_\mu$. The small first-order spin-rotation coupling effect (mentioned in Sec.~\ref{sec:perturbation}, Eq.~\ref{eq:SR1}) will be evaluated on the phase space electronic states and added as a correction to the phase space energy splitting.
To predict the spin-rotation splitting associated with rotations of a free molecule in 3D, one can combine the spin-rotation coupling tensor from a PS calculation with the 3D spin-rotation model in Sec.~\ref{sec:perturbation}; for example, for the symmetric CH3 molecule, we used Eq.~\ref{eq:Esr_NK} to generate the data in Table~\ref{table:peak}.

\section{\textit{Ab Initio} Results and Discussions} \label{sec:results}
The theory above was implemented for a set of small molecules. All computational details are available in the SI. Importantly, note that experimentally, one first extracts energy splittings and then second fits spin rotation tensors. Theoretically, however, one initially calculates the spin-rotation tensor and then, if one wishes, one can predict energy splittings. To that end, we will mostly compare tensors, but we will also explicitly compare energy splittings for one molecule (CH$_3$) as well in order to highlight the power of a PS electronic structure approach. Because the experiments in Refs.~\citenum{yamada1981diode,davis1997jet,amano1982difference,endo1982microwave,tanimoto1999microwave,schuder_sub-doppler_2017} report only the diagonal elements of $\epsilon$, we will not compute off-diagonal matrix elements below. 

\subsection{CH$_3$} \label{sec:CH3}
CH$_3$ is a planar symmetric top radical with an unpaired electron and $C_3$ rotational symmetry. Under the relativistic BO approximation, the ground electronic states are a degenerate Kramers doublet with the unpaired electron lying primarily in the $p_c$ orbital where $c$ is the $C_3$ symmetry axis. Following the intuition from Sec.~\ref{sec:ps_SR}, a nuclear rotation around the  $\mu$-axis, $\mu=a,b,c$, effectively generates an electronic current around $\mu$ with finite $L_\mu$, which then couples with SOC to form spin-rotation coupling. Due to the symmetry of CH$_3$, the spin-rotation coupling tensor is diagonal in the principal axes, and the diagonal elements are degenerate between the $a$- and $b$-axes. Because the density of the unpaired electron lies mostly along the $c$-axis, rotations around an in-plane axis ($a$ or $b$) generate a much stronger electric current and therefore stronger spin-rotation coupling, so that $\epsilon_a = \epsilon_b > \epsilon_c$. This trend is evident from the spin-rotation coupling constant fitted to experimental rotational spectra: $-354$ MHz for rotation around the $a$ and $b$ axes and only about $-3$ MHz for rotation around the $c$ axis~\cite{davis1997jet,roberts2012high,amano1982difference,yamada1981diode,roberts2012sub}, i.e., about 2 orders of magnitude in difference.

\begin{table}[htb!]   
  \centering
  \begin{tabular}{c c c c c}
      \toprule
           Molecule & rotation axis & experiment (MHz) & theory (MHz) & error \\ \midrule
           \multirow{2}{*}{CH$_3$} & a/b & $-$354(5)~\cite{davis1997jet} & $-$342 & 3.3\%  \\ 
                                   & c   & $-$3(63)~\cite{davis1997jet,roberts2012sub,amano1982difference,yamada1981diode}  & $-$7 & within uncertainty\\ \cmidrule(lr){1-5}
           \multirow{2}{*}{CF$_3$} & a/b & $-$36.500(42)~\cite{endo1982microwave} & $-$41.097 & 12.6\%  \\
                                   & c   & 3.35(15)~\cite{endo1982microwave}  & 3.21 & 4.2\% \\ \cmidrule(lr){1-5}
           \multirow{2}{*}{SiF$_3$} & a/b & 36.015(33)~\cite{tanimoto1999microwave} & 33.599 & 6.7\%  \\
                                   & c   & 4.810(22)\cite{tanimoto1999microwave}  & 4.512 & 6.2\% \\ \cmidrule(lr){1-5}
           \multirow{3}{*}{CH$_2$OH}& a   & $-$512.5(17)\cite{schuder_sub-doppler_2017} & $-$467.5 & 8.8\%  \\
                                   & b   & $-$125.2(3)\cite{schuder_sub-doppler_2017}  & $-$117.9 & 5.8\% \\                               
                                   & c   & $-$3.7(3)\cite{schuder_sub-doppler_2017}  & 0.2 & 106.6\% \\                               
              \bottomrule 
  \end{tabular}  
  \caption{Predicted spin-rotation coupling constants of CH$_3$, CF$_3$, and CH$_2$OH in comparison to experiments. The numbers in brackets under ``experiment'' are the experimental uncertainties. The error is defined as $|(\Delta\nu_\mathrm{exp}-\Delta\nu_\mathrm{theory})/\Delta\nu_\mathrm{exp}|$. }
  \label{table:coupling}
\end{table}

Calculations for the spin-rotation coupling constants of CH$_3$ are shown in Table~\ref{table:coupling}. For the larger coupling constant, associated with rotation around $a$ and $b$, our prediction is $-342$ MHz, with only 3.3\% deviation from the experimental measurement. The other coupling constant is very small and was reported with a large experimental uncertainty of 63 MHz~\cite{yamada1981diode}. Our PS method predicts $-7$ MHz and is well within the experimental uncertainty range. Our prediction is also comparable to and has a slightly smaller error than the predictions using response calculations and the coupled cluster singles and doubles (CCSD) solver at the same basis: $|\epsilon_{bb}|=399$ MHz, $|\epsilon_{cc}|=3$ MHz, corresponding to an overall 10.6\% error~\cite{tarczay2010first}.

\begin{figure}[htb!] 
    \centering
    \includegraphics[width=0.5\columnwidth]{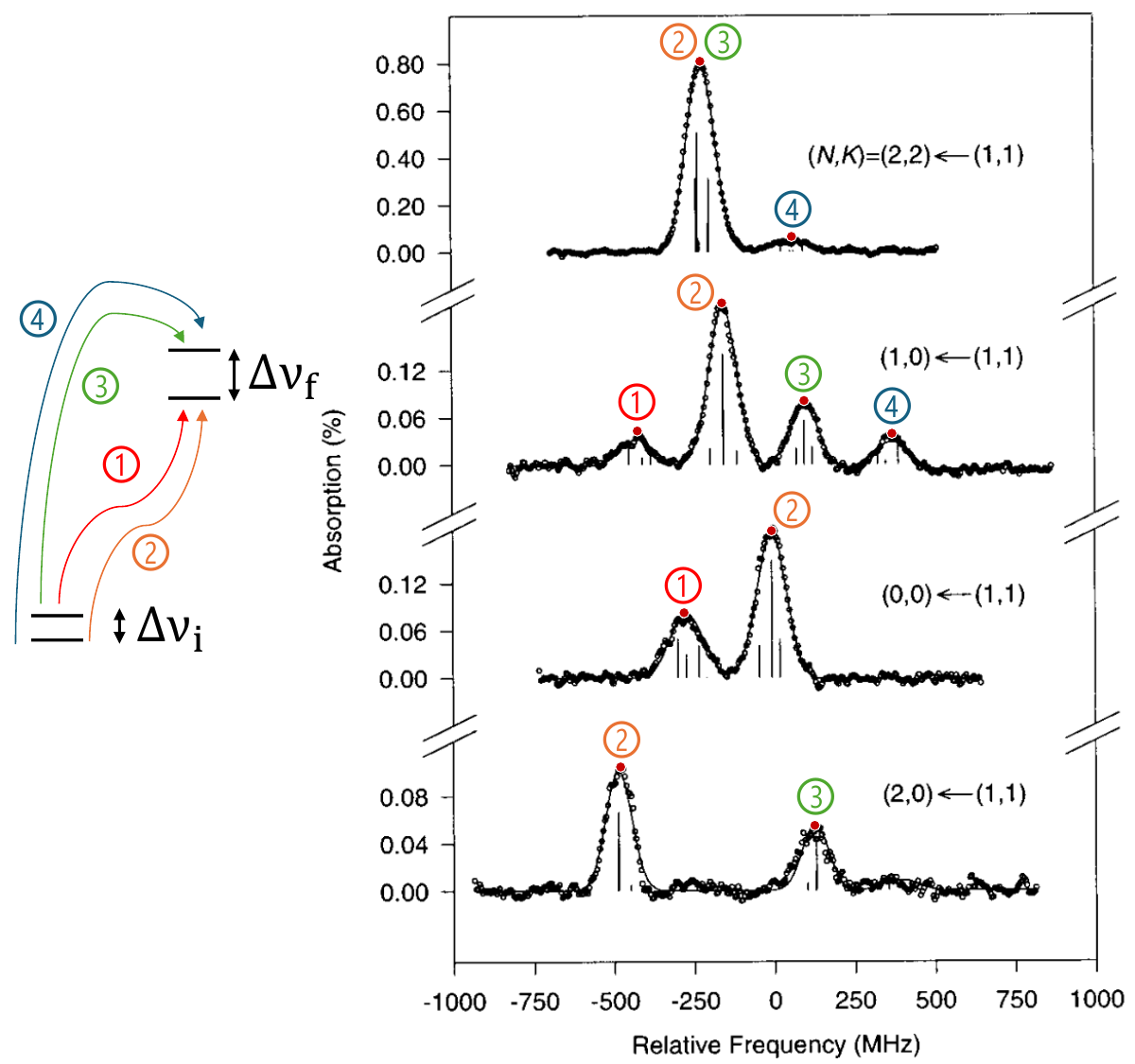}
    \caption{Spin-rotation coupling splitting of CH$_3$ observed in the rotational spectroscopy. Each peak is labeled based on the type of transition; on the left hand side, $\Delta\nu_i$ and $\Delta\nu_f$ represent the spin-rotation splitting of the initial and final states, respectively. The original spectrum is reproduced from Ref.~\citenum{davis1997jet} with permission of AIP publishing. The frequencies between the split levels are plotted relative to the frequency difference between the unsplit transitions (according to an experimental fit). The experimental transitions here are from vibration level $v=0$ to vibrational level $v=1$, but changes in vibrational level are removed by considering the relative frequency.}
    \label{figure:nesbitt_CH3}
\end{figure}

\begin{table}[htb!]   
  \centering
  \begin{tabular}{l c c c}
      \toprule
            Splitting & experiment (MHz) & theory (MHz) & error \\ \midrule
            \multicolumn{4}{c}{$(N,K) = (2,2) \leftarrow (1,1)$\rule[-1.5ex]{0pt}{2ex} } \\
           $\circled{4}\leftrightarrow(\circled{3} + \circled{2})/2$ & 288 & 279  & 3\%  \\ \cmidrule(lr){1-4}
            \multicolumn{4}{c}{$(N,K) = (1,0) \leftarrow (1,1)$\rule[-1.5ex]{0pt}{2ex}} \\ 
           $\circled{2} \leftrightarrow \circled{1}$ & 262 & 262  & 0\%  \\
           $\circled{3} \leftrightarrow \circled{2}$ & 256 & 252  & 1\%  \\
           $\circled{4} \leftrightarrow \circled{3}$ & 274 & 262  & 5\%  \\  \cmidrule(lr){1-4}
            \multicolumn{4}{c}{$(N,K) = (0,0) \leftarrow (1,1)$\rule[-1.5ex]{0pt}{2ex}} \\
           $\circled{2} \leftrightarrow \circled{1}$ & 274 & 262  & 5\%  \\  \cmidrule(lr){1-4}
            \multicolumn{4}{c}{$(N,K) = (2,0) \leftarrow (1,1)$\rule[-1.5ex]{0pt}{2ex}} \\
           $\circled{3} \leftrightarrow \circled{2}$ & 615 & 594  & 3\%  \\ 
           \bottomrule 
  \end{tabular}  
  \caption{The splitting of the different peaks in the rotational spectra of CH$_3$ due to spin-rotation coupling. See Fig.~\ref{figure:nesbitt_CH3} for peak labeling.  The error is defined as $|(\Delta\nu_\mathrm{exp}-\Delta\nu_\mathrm{theory})/\Delta\nu_\mathrm{exp}|$. The experimental values were extracted from Ref.~\cite{davis1997jet} with WebPlotDigitizer\cite{WebPlotDigitizer}. }
  \label{table:peak}
\end{table}

Given the accurate prediction of the spin-rotation coupling constants, each associated with rotations around a single molecular axis, we can further predict the couplings for the rotations of the free molecule in 3D. Because CH$_3$ is a symmetric top molecule, we will evaluate the spin-rotation splitting of each rotational level $(N,K)$ using the analytical expression Eq.~\ref{eq:Esr_NK}. This calculation enables direct comparison with the relative frequency of individual peaks in the experimental rotational spectra. Each peak in the spectra corresponds to a transition between a spin-rotation split state of the initial rotational level and a spin-rotation split state of the final rotational level. In Fig.~\ref{figure:nesbitt_CH3} on the left, we have classified each transition into four types, ordered by the transition frequency. On the right, we label each peak based on the transition type and compare with the theoretical prediction of the associated splitting.
The tall, broad signal in the top spectrum for the transition from $(N,K)=(1,1)$ to $(N,K)=(2,2)$ is formed by two closely spaced transitions, so the peak frequency is compared to the average of these two transitions. In Table~\ref{table:peak}, we tabulate the splitting between any two adjacent peaks and compare the experimentally found splitting with the theoretical prediction. On average, the relative error of the theory is only 3\%.
This accuracy is especially impressive after taking into account the experimental noise. Even the reported $\epsilon$ directly fitted to the experimental splitting predicts splitting of on average 2\% error. The agreement between the PS prediction and the experimental spectra is a strong validation of the PS method as a means to capture spin-rotation coupling quantitatively. 

\subsection{CF$_3$ and SiF$_3$}
CF$_3$ and SiF$_3$ are two symmetric top radicals known for having major spin-rotation coupling constants $\epsilon_{aa}$ of similar magnitude but opposite sign. In this section, we apply our PS method to these two systems and demonstrate that PS PESs can correctly capture not only the magnitude but also the sign of the spin-rotation coupling constants. 

The sign of the spin-rotation coupling constant is determined by the energy ordering between the $J=N+S$ and $J=N-S$ states, and in the case of classical nuclear rotation, simplifies to the preferred alignment between $\mathbf{N}$ and $\mathbf{S}$. If $N_\mu$ and $S_\mu$, $\mu=a,b,c$, are aligned in the lower spin-rotation split state and antialigned in the higher energy state, then $\epsilon_\mu$ is negative, representing an energy stabilization when $N_\mu$ and $S_\mu$ align; vice versa, antialigned $N_\mu$ and $S_\mu$ in the lower state implies a positive $\epsilon_\mu$.
At a small positive $N_a$, the lower PS electronic state of CF$_3$ has $\mathbf{S}\approx(0.5,0,0)$. This $\mathbf{S}$ aligns with $\mathbf{N}$ and thus suggests a negative $\epsilon_{aa}$. On the contrary, the lower PS electronic state of SiF$_3$ has $\mathbf{S}\approx(-0.5,0,0)$, which suggests a positive $\epsilon_{aa}$ for SiF$_3$. As shown in Table~\ref{table:coupling}, these signs predicted by the PS method indeed agree with the signs measured by experiments. The PS electronic states further help understand what controls the sign: the SOC between the ground and excited state. In either molecule, the electronic orbital angular momentum $\mathbf{L}$ of the lower PS state aligns with $\mathbf{N}$ in the lower energy states of both molecules, which follows because the Coriolis Hamiltonian term $-\mathbf{I}^{-1}\mathbf{N}\cdot \mathbf{L}$ contains a minus sign. Therefore, the key question is the direction of $\mathbf{S}$, and the relative orientation between $\mathbf{S}$ and $\mathbf{L}$ is determined by the SOC between ground and excited states.
More quantitatively, from the data in Table~\ref{table:coupling}, the magnitude of the spin-rotation coupling constants, calculated in the same way as discussed in Sec.~\ref{sec:CH3}, are also accurate compared to the experimental best fit, with relative errors around 10\%. 

\subsection{CH$_2$OH}
Finally, we study a molecule with lower symmetry, CH$_2$OH. As an asymmetric top, its spin-rotation coupling tensor is not diagonal and has three nondegenerate diagonal elements. The large number of unique parameters (in general, 9) in the spin-rotation tensor makes it difficult to recover the full tensor from purely experimental means. As a result, in many experimental analyses, only the diagonal elements $\epsilon_{\mu\mu}$ are fitted to the spectra, and the off-diagonal elements are approximated to be 0.
Thus, we report here only the diagonal elements; we provide our predicted off-diagonal elements in the SI Sec.~\ref{sec:supp_ch2oh}. 

Compared to experimentally fitted diagonal elements in Table~\ref{table:coupling}, the two larger diagonal spin-rotation coupling constants, $\epsilon_a$ and $\epsilon_b$ are predicted accurately by PS with less than 10\% difference. The smallest coupling constant, $\epsilon_c$ was predicted to be near zero because the second-order spin-rotation splitting captured by PS is largely cancelled by the first-order term (Eq.~\ref{eq:SR1}) of a similar magnitude. Whereas the first-order spin-rotation splitting is not very sensitive to basis and DFT functional, the second-order splitting varies significantly with the functional and thus is difficult to predict accurately without a correlated wavefunction method. For the same reason, the sign of the resulting coupling constant depends heavily on the DFT functional. Thus, calculating this element is a challenge: the difference in the sign of $\epsilon_{cc}$ between our prediction and the experimentally fitted parameter has also been observed in the previous theoretical prediction using B3LYP, which predicted $\epsilon_{cc}=9.3$ MHz.

\section{Conclusion: Kramers' Degeneracy and a Conceptual Shift}~\label{sec:conclusion}
We have demonstrated that, for a variety of small radical molecules,  high-resolution spin-rotation couplings can be extracted naturally when we give up on the notion of Born-Oppenheimer potential surfaces and instead run electronic structure calculations using PS electronic structure theory (with energies and molecular orbitals that depend on both nuclear position $\mathbf{X}$ and nuclear momentum $\mathbf{P}$). 
Our predictions are quantitatively accurate compared to experiments, not only for the magnitude but also for the sign of the coupling constant, for both symmetric top systems and systems with lower symmetry. We note that the present paper follows a previous paper on $\Lambda$-doubling\cite{peng2026phasespaceelectronicstructure},  where we also found quantitative accuracy for $\Lambda$-splittings of the NO molecule using PS electronic structure theory. For both spin-rotation couplings and $\Lambda$-doubling calculations, PS electronic structure theory avoids any and all summations over intermediate states; mathematically, relative to BO theory,  one gains a non-perturbative means to evaluate the resolvent in a stable fashion with a rich set of new physics to be learned.

One of the most impactful lessons made clear by running PS electronic structure theory is that, for many years now, there has been confusion about eigenvalue degeneracy within the chemical community. According to Kramer's theorem, because CH$_3$ contains an odd number of electrons, its eigenvalues must be (at least) doubly degenerate.  That being said, however, it is crucial to recognize that Kramers' theorem really has two incarnations. $(i)$ On the one hand,  one can apply Kramers' theorem to the BO electronic structure calculation for a radical molecule (like CH$_3$). One concludes that all BO potential energy surfaces must be doubly degenerate -- a fact well known to all chemists.  (ii) On the other hand, experimentalists do {\em not} observe the eigenvalues of an electronic problem; instead, they observe the eigenvalues of the total nuclear plus electronic plus spin Hamiltonian.  To that end,  one can also apply Kramers' theorem to the total Hamiltonian, and one must still recover (at least) double degeneracy. How should we reconcile this necessary degeneracy with the existence of spin-rotation coupling which splits the spin states in Fig.~\ref{fig:PES}? Does one break Kramers' theorem? More generally, how should we understand degeneracy in Fig.~\ref{fig:PES}?

To that end, it is crucial to emphasize that the total Kramers' degeneracy for a CH$_3$ molecular eigenstate arises {\em not exclusively from the degeneracy of the BO electronic levels} but also from the rotations of the radical molecule relative to a fixed lab frame axis.  In other words, within the BO framework, one must (obviously) write down the full wavefunction (with electronic + vibrational + rotation components) and then make an argument about which states are degenerate and perhaps apply the time-reversal operator. In the context of PS theory, this approach is much more nuanced because when one explores the $\mathbf{P}$ direction, one is already accounting for vibrations and rotations within a classical framework and, as such, reading off degeneracy from Figs.~\ref{fig:PES}b and \ref{fig:PES}c is not straightforward. Moreover, the quantization of angular momentum for a rigid body requires 3-dimensional boundary conditions which cannot be ascertained along a single reaction coordinate; effectively, the $x$-axis in Figs.~\ref{fig:PES}b and \ref{fig:PES}c represents the projection of angular momentum along a given direction (not the total angular momentum).

Consider  CH3 as an example.  Quantum mechanically, using a BO framework, we know 
that because the CH3 ground state is roughly of the form $\ket{X; N=0, K=0, S=1/2, \Sigma = \pm 1/2}$ where $X$ represents the electronic ground states and $\Sigma$ represents the projection of the electron spin on the molecular symmetry axis. For this problem, the ground state degeneracy is clearly twofold and arises from the spin-degeneracy; within a PS framework, this ground state is represented by the two states crossing at $L^\mathrm{n}=0$ in Fig.~\ref{fig:PES}.
That being said, the first and second excited states are more complicated.  If we consider CH3 to be an oblate symmetric top with $N$ and $K$ reasonably good quantum numbers, we expect the body-frame eigenstates of interest to be the two states $\ket{X; N=1, K=1}$ and 
$\ket{X; N=1, K=-1}$.
(If we were to consider the CH$_2$OH molecule without any symmetry, then the first excited rotational state would be entirely non-degenerate; here, $K$ would not be a good quantum number.)  Next, we must consider that the relevant excited states can rotate not only about the molecular axis, but also through rotation of the molecular axis in the space, which leads to a $2N+1 = 3$ extra degeneracy associated with the space-frame projection quantum number $M_N$. In other words, for CH3 we expect the first and second excited states to emerge from 6 degenerate rotational states; for CH$_2$OH, the first and second excited states should emerge from 3 degenerate rotational states. 
Finally, if we further couple to spin, and work with the states $\ket{X; N=1, K=1, S=1/2, \Sigma = \pm 1/2}$ and $
\ket{X; N=1, K=-1, S=1/2, \Sigma = \pm 1/2}$, there are now 12 low-lying, degenerate or nearly degenerate excited states for CH3 from which to choose the first and second excited states (and 6 degenerate or nearly degenerate states from which to choose
the first and second excited states for CH$_2$OH).  

Now, it is crucial to emphasize that once we couple the $N=1$ nuclear rotations to the spin, we can decompose the resultant basis functions into doublet ($J$=1/2) and quartets ($J$=3/2). In particular, for CH3 we expect two quartets and two doublets to emerge for the low-lying excited states (and one quartet and one doublet for CH$_2$OH).   All of this information is hidden in Fig.~\ref{fig:PES}. Here, if we consider $L^n$ to represent $K$,  one certainly cannot discern the degeneracy of the given states along the $x$-axis.  In fact, from the discussion above, we know that we should expect a doublet and a quartet for the lower and upper curves in Fig.~\ref{fig:PES}, but which one is which is not obvious and depends on the sign of the spin-rotation coupling constant. 

Thus, at the end of the day, whereas each PS surface represents a single state at $L^\mathrm{n}= 0$ in Fig.~\ref{fig:PES}c, each surface can represent multiple quantum states at $L^\mathrm{n} = \pm \hbar$. 
On the one hand, this uncertainty in degeneracy must make a chemist wary and would appear to be one limitation of using a semiclassical (rather than quantum) view of angular momentum; on the other hand, we note that the entire discussion of degeneracy and rotational coupling is missing within BO theory and PS clearly offers the chemist the ability to measure many more experimental observables than the existing BO framework; thus a more nuanced view of degeneracy may be necessary within a practical scheme.
(See also Ref. \citenum{peng2026phasespaceelectronicstructure} for a mathematical discussion of how to interpret Kramers' theorem for $\Lambda$-doubling in the case of the NO molecule.)
Ultimately, when one runs a PS electronic structure calculation, one is immediately reminded that the eigenstates of the BO Hamiltonians are not really
``stationary,''  that corrections for nuclear motion are essential,
and that one must be very careful when applying Kramers' theorem to the electronic structure problem with frozen nuclei and then naively extrapolating the conclusions to quantities that can be measured experimentally.  Within this lens, because the nuclei are always moving, any meaningful calculation of electronic structure must properly be understood as a calculation in a non-inertial time-dependent frame -- where the electronic Hamiltonian is not invariant to time-reversal and where  Kramers' theorem does not directly apply.

Looking ahead, one wonders exactly how much there is still to learn when we switch from BO calculations to PS calculations. Spin-rotation coupling and $\Lambda$-splitting have already been addressed;
the  Einstein-de Haas\cite{einstein1915experimenteller,mentink2019quantum, ganzhorn2016quantum} and  Barnett\cite{barnett1915magnetization} effects seem well within reach.   Most importantly, PS electronic structure also has the potential to treat far more complicated and non-perturbative many-body phenomena far out of reach of BO theory, including
the emergent chiral induced spin selectivity\cite{bloom2024chiral, ray1999asymmetric, gohler2011spin} (CISS) and perhaps even superconductivity\cite{bardeen1957theory,tinkham2004introduction}. 
For all of these reasons, we contend that moving away from Born-Oppenheimer theory and running PS electronic structure theory can offer an important shift in our understanding of molecules and materials with degenerate or nearly degenerate ground electronic states.

\section{Computational details}

All calculations were performed using a phase space (PS) implementation based on the PySCF package~\cite{sun2018pyscf,sun2020recent}. The PS Hamiltonian includes the $\mathbf{P}\cdot \hat{\mathbf{\Gamma}}$ electron-nuclear coupling term and not the $\hat{\mathbf{\Gamma}}^2$ term because $\langle\hat{\mathbf{\Gamma}}^2\rangle$ is orders of magnitude smaller than the coupling term and shifts the two PESs near equally when $\hat{\mathbf{\Gamma}}_A = \hat{\mathbf{\Gamma}}_A^\mathrm{spatial}$ as considered in this work. We used the Breit-Pauli Hamiltonian~\cite{bethe2013quantum} for both the 1-electron and 2-electron spin-orbit coupling. For the 2-electron SOC, we adopted the spin-orbit mean-field (SOMF) approximation~\cite{HebetaSOMF1996}. Because calculations of spin-rotation coupling can be relatively sensitive to the electronic basis~\cite{tarczay2010first}, we solved the PS problems in large bases (aug-cc-pVTZ and cc-pVQZ~\cite{dunning1989gaussian,woon1993gaussian}) that numerically converge the spin-rotation energy splitting. To treat the large number of electronic orbitals efficiently, the electronic Hamiltonian was solved using the generalized Kohn-Sham (GKS) method with the TPSS functional~\cite{tao2003climbing} and with the multicollinear extension ``mcfun''\cite{pu2023noncollinear} to handle noncollinear spin. See Sec.~\ref{sec:supp_DFT} for more details on benchmarking DFT functionals and orbital bases. When calculating the spin-rotation energy splitting associated with the diagonal spin-rotation coupling constant $\epsilon_{\mu\mu}$, $\mu=a,b,c$, we performed one GKS calculation at $L^\mathrm{n}_\mu =\hbar$ with an initial guess of $\mathbf{S}$ aligned with the $\mu$ axis ($S_\mu\approx1/2$), and then used the time reversal of the converged GKS solution as the initial guess for a second GKS calculation at $L^\mathrm{n}_\mu =\hbar$ to obtain the other GKS minimum with $S_\mu\approx-1/2$. The energy difference of these two GKS solutions is defined to be $\Delta E^\mathrm{SR}$, which was used to fit $\epsilon_{\mu\mu}$.

\section*{Acknowledgement}
We acknowledge helpful discussions with Melanie Ann Roberts Reber.  We dedicate this article to our friend, David Waldeck, a true pioneer of spin interactions, who passed away far too early. 

\putbib
\end{bibunit}



\begin{bibunit}
\clearpage

\begin{center}
\textbf{\Large Supporting information for: \\ A Conceptual Shift In Our Understanding of Degenerate Radical Spin Systems: Spin-Rotation Coupling Turned On Its Head} \\

\vspace{1em}
\end{center}

\setcounter{equation}{0}
\setcounter{figure}{0}
\setcounter{table}{0}
\setcounter{page}{1}
\setcounter{section}{0}
\setcounter{secnumdepth}{3}
\renewcommand{\theequation}{S\arabic{equation}}
\renewcommand{\thefigure}{S\arabic{figure}}
\renewcommand{\thetable}{S\arabic{table}}
\renewcommand{\thesection}{S\arabic{section}}

\section{DFT functional and basis dependence} \label{sec:supp_DFT}
\begin{figure}[htb!]
    \centering
    \includegraphics[width=0.5\columnwidth]{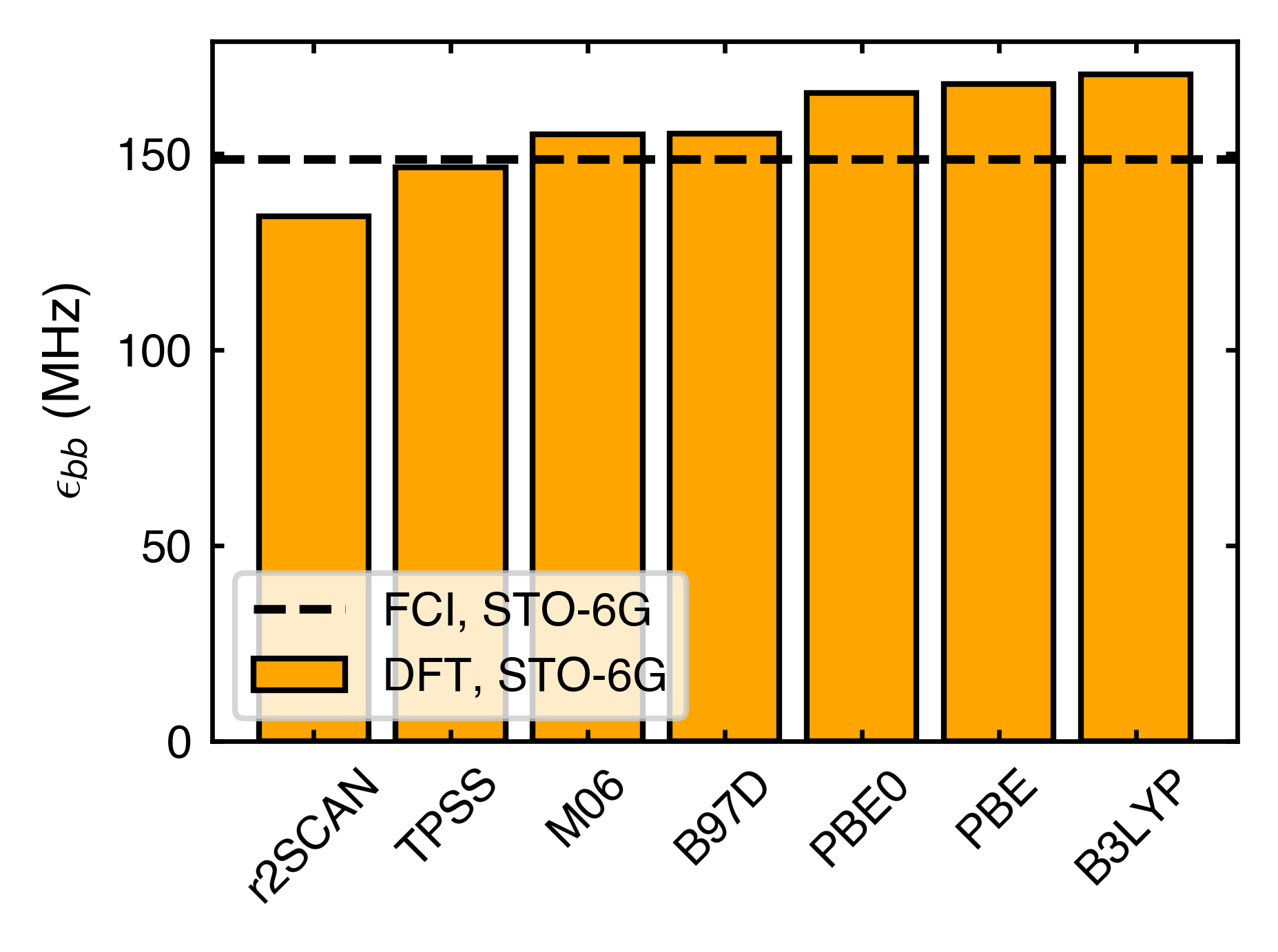}
    \caption{Predicted spin-rotation coupling tensor element $\epsilon_{bb}$ of CH$_3$ using various DFT functionals in comparison to the exact solver FCI in the minimal basis (STO-6G).}
    \label{fig:functional}
\end{figure} 

\begin{figure}[htb!]
    \centering
    \includegraphics[width=0.5\columnwidth]{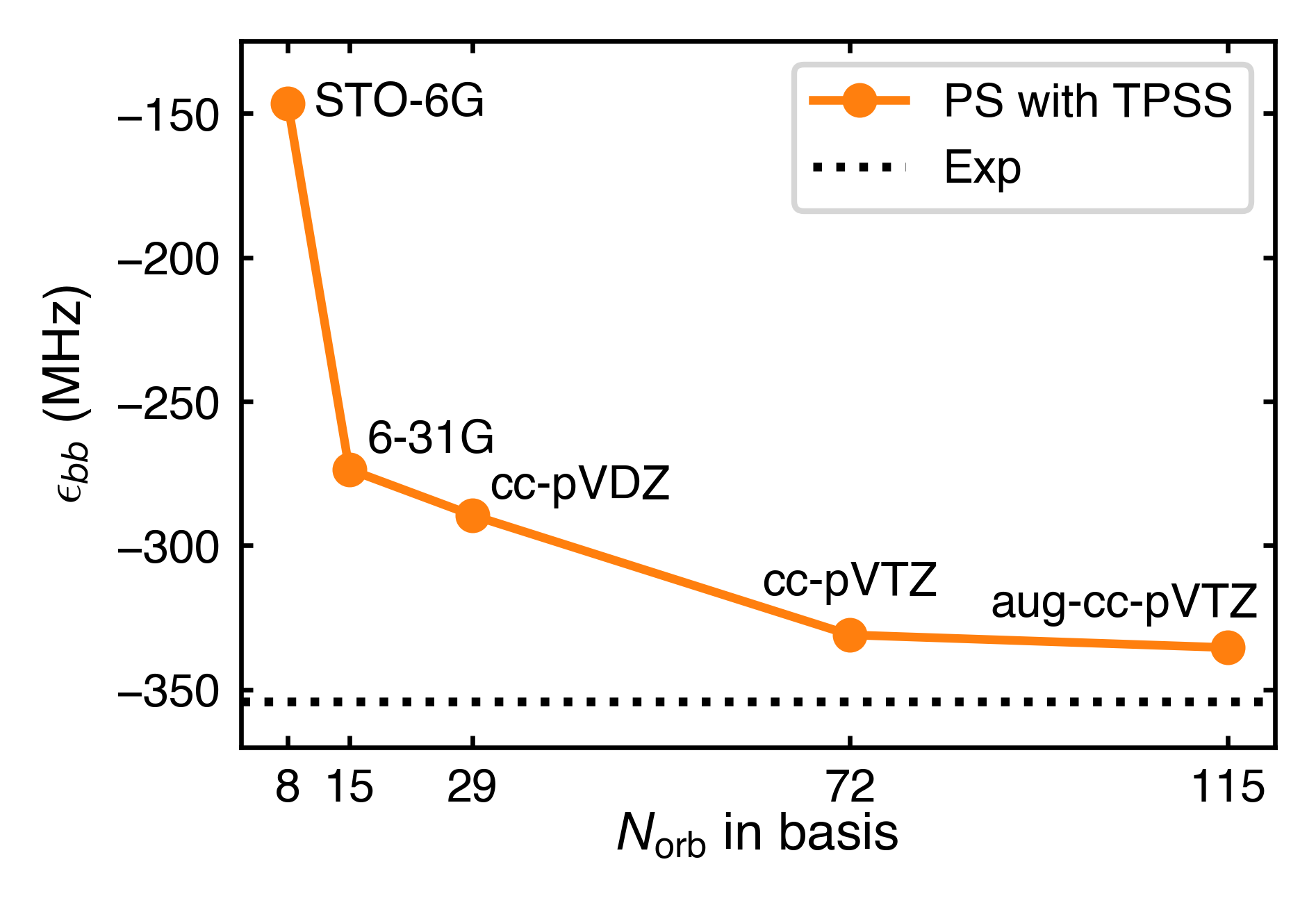}
    \caption{Predicted spin-rotation coupling tensor element $\epsilon_{bb}$ of CH$_3$ in various bases with TPSS functional. The black dashed line is the $\epsilon_{bb}$ extracted from experimental spectra~\cite{davis1997jet}.}
    \label{fig:basis}
\end{figure}
The generalized Kohn-Sham (GKS) energies obtained with different DFT functionals were benchmarked against full configuration interaction (FCI) and density matrix renormalization group (DMRG)~\cite{chan2011density} calculations for small basis sets for the CH$_3$ molecule.  In the minimal STO-6G basis, as shown in Fig.~\ref{fig:functional}, the TPSS functional~\cite{tao2003climbing} yields an energy splitting in closest agreement to FCI. Fig.~\ref{fig:functional} also shows that the spin-rotation coupling prediction varies only slightly with the DFT functional choice. This insensitivity suggests that the PS electronic Hamiltonian is numerically stable. We further benchmarked TPSS in a larger basis, 6-31G, against DMRG implemented in block2~\cite{zhai2023block2}. Here, the DMRG benchmark calculations used a tight Davison convergence threshold of $10^{-8}$ and sufficiently large bond dimensions (up to 2000) to reduce the discarded weight to  $8\times10^{-8}$, thereby providing a near-exact reference. In 6-31G basis, compared to DMRG, TPSS underestimated the spin-rotation coupling splitting by about 7.3\%. We further tested basis set convergence using the TPSS functional (Fig.~\ref{fig:basis}). At aug-ccpVTZ, the difference in the predicted $\epsilon_{bb}$ compared to ccpVTZ was reduced to 1\%. For this reason, for all studies in the main text, we ran TPSS calculations in the aug-ccpVTZ basis except for SiF$_3$, where a ccpVQZ basis was used. 

\section{More details about Eq.~\ref{Eq:E2_PT}}

According to Eq.~\ref{Eq:E2_PT}, one can generate the matrix element $\epsilon_{ab}$ in three steps. First, for a doubly degenerate ground state for the $\hat{H}_0$ Hamiltonian $\{ \ket{0}, T\ket{0}\}$, one must find the coefficients $c_1$ and $c_2$ such that the vector $\ket{0_a} \equiv c_1 \ket{0} +c_2 T\ket{0} $ has spin expectation value

$$ \left< 0_a \middle | \hat{\mathbf{S}} \cdot \mathbf{e}_a \middle | 0_a \right> = \frac{\hbar}{2} $$

Second, if we imagine including $H_\mathrm{SOC}$, we apply first order perturbation theory to $\ket{0_a}$ and generate:

$$\ket{\Phi_{\mathrm{SOC},a}^{(1)}} = \ket{0_a}  +  \sum_{i\neq 0} \frac{2|i\rangle \langle i|\hat{H}_\mathrm{SOC}|0_a\rangle}{E_{0} - E_i} $$
This first-order wavefunction correction is the key element in the second-order energy expression for spin-rotation coupling
\begin{equation}
    E_a^\mathrm{SR(2)} = 2\langle 0 | -\mathbf{I}^{-1} \mathbf{N}   \cdot\hat{\mathbf{L}}|\Phi_{\mathrm{SOC},a}^{(1)}\rangle \approx -\mathbf{I}^{-1} \langle \Phi_{\mathrm{SOC},a}^{(1)} | \mathbf{N} \cdot \hat{\mathbf{L}}|\Phi_{\mathrm{SOC},a}^{(1)}\rangle
\end{equation}

Third and finally, for one quantum of nuclear rotation around each axis $\mathbf{N}=\mathbf{e}_\mu$, $\mu=a,b,c $, we evaluate the associated angular momentum expectation value for state $\ket{\Phi_{\mathrm{SOC},a}^{(1)}}$
\begin{equation} \label{eq:L_vec}
    L^a_\mu =  \left< \Phi_{\mathrm{SOC},a}^{(1)} \middle | \hat{\mathbf{L}} \cdot \mathbf{e}_\mu\middle|
\Phi_{\mathrm{SOC},a}^{(1)} \right>
\end{equation}
and set the spin-rotation coupling tensor 
$$\epsilon_{ab} =  - 2\sum_\mu[\mathbf{I}^{-1}]_{b \mu}  L^a_\mu.$$

Within a density matrix formalism, this procedure is equivalent to generating a $2 \times 2$ matrix $M$
\begin{equation}
    M  = 
    \left(
    \begin{array}{cc}
    \displaystyle \sum_{i\neq 0} \frac{2\langle 0|\hat{H}_\mathrm{SOC}|i\rangle \langle i|-\mathbf{I}^{-1} \mathbf{N} \cdot\hat{\mathbf{L}}|0\rangle}{E_{0} - E_i} 
    &
     \displaystyle\sum_{i\neq 0} \frac{2\langle 0|\hat{H}_\mathrm{SOC}|i\rangle \langle i|-\mathbf{I}^{-1} \mathbf{N} \cdot\hat{\mathbf{L}}|0'\rangle}{E_{0} - E_i} 
        \\
     \displaystyle\sum_{i\neq 0} \frac{2\langle 0'|\hat{H}_\mathrm{SOC}|i\rangle \langle i|-\mathbf{I}^{-1} \mathbf{N} \cdot\hat{\mathbf{L}}|0\rangle}{E_{0} - E_i} 
    &
    \displaystyle\sum_{i\neq 0} \frac{2\langle 0'|\hat{H}_\mathrm{SOC}|i\rangle \langle i|-\mathbf{I}^{-1} \mathbf{N} \cdot\hat{\mathbf{L}}|0'\rangle}{E_{0} - E_i} 
    \end{array}
    \right)
\end{equation} 
where $\ket{0}$ are $\ket{0'}=T\ket{0}$ are elements of the degenerate ground state basis. In order to evaluate $\epsilon_{ab}$, one then must find a spin density matrix $\rho$ such that 

$$
\mathrm{Tr}((\hat{\mathbf{S}} \cdot \mathbf{e}_a) \rho) = \frac{\hbar}{2}
$$
and then evaluate
$$
\epsilon_{ab} \equiv \frac{\partial \mathrm{Tr} \left( \rho M \right)}{\partial N_b}
$$

These procedures are not very efficient as they require a sum over states when generating Eq.~\ref{Eq:E2_PT}, and thus the second equality in Eq.~\ref{eq:CC_SR} is usually used instead. In the next section, we will remind the reader that $L_\mu^a$ in Eq.~\ref{eq:L_vec} is closely related to the electronic $g$ tensor as present in a magnetic field, so that one can also approximate the $\epsilon$ tensor using the $g$ tensor (which is often experimentally easier to evaluate).

\section{Standard calculation of spin-rotation coupling tensor from $g$-tensor anisotropy}
Consider the electronic Hamiltonian
\begin{equation}
    \hat{H} = \hat{H}_0 + \hat{H}_\mathrm{SOC} + \hat{H}_\mathrm{Zeeman}
\end{equation}
where $\hat{H}_0$ is the nonrelativistic BO Hamiltonian, the spin-orbit coupling $\hat{H}_\mathrm{SOC} = A \mathbf{\hat{S}} \cdot \mathbf{\hat{L}}$ and the Zeeman term 
\begin{equation} \label{eq:zeeman}
    \hat{H}_\mathrm{Zeeman} = \mu_B \mathbf{B}\cdot (\mathbf{\hat{L}} + g_e \mathbf{\hat{S}})
\end{equation}
are relatively small and can be treated perturbatively. In many studies on magnetic effects, one seeks to approximate the Zeeman term with an effective Hamiltonian
\begin{equation} \label{eq:eff}
    \hat{H}_\mathrm{eff}^{\mathrm{Zeeman}} = \mu_B  \mathbf{B} \cdot g_\mathrm{eff} \cdot \mathbf{\hat{S}}
\end{equation}
where $g_\mathrm{eff}$ is a fitted effective $g$-tensor designed to make $\hat{H}_\mathrm{eff}^{\mathrm{Zeeman}}$ match $\hat{H}_\mathrm{Zeeman}$ within the two-dimensional ground state subspace  $\left\{ |0\rangle, |0'\rangle \right\}$ for the nonrelativistic $H_0$. In many systems such as CH$_3$, the electronic orbital angular momentum is quenched under the nonrelativistic BO approximation~\cite{atkins2011molecular}, i.e. $\langle0|\mathbf{\hat{L}}|0\rangle = \langle0'|\mathbf{\hat{L}}|0'\rangle = 
\langle0|\mathbf{\hat{L}}|0'\rangle = \mathbf{0}$. Thus, to first order, the effective $g$-factor $g_\mathrm{eff}$ is the same as the isotropic $g_e\approx2.0023$. However, if we go to second order, then the SOC will induce a nonzero $\langle\mathbf{\hat{L}}\rangle$ such that
$g_\mathrm{eff}$ must be adjusted in an anisotropic fashion.
Below, we will outline how this $g$-factor anisotropy (more precisely, deviation of $g$-factor from $g_e$) is obtained.

Assuming a BO ground state can be written in terms of a product of a spatial part and a spin part $|0, \sigma\rangle = |0 \rangle |\sigma\rangle$, the first-order corrected wavefunction due to SOC is 
\begin{align}
    |0^{(1)}\rangle=& |0, \sigma\rangle + \sum_{i \ge 1,\sigma'} |i,\sigma' \rangle \frac{\langle i, \sigma'| A \mathbf{\hat{S}} \cdot \mathbf{\hat{L}} | 0, \sigma \rangle}{E_0 - E_{i, \sigma'}}
\end{align}
which gives rise to a nonzero $\langle\mathbf{\hat{L}}\rangle$,
\begin{align}
    \langle 0^{(1)}|  \mathbf{\hat{L}} | 0^{(1)} \rangle \approx & 2 \sum_{i\ge 1,\sigma'}  \frac{\langle 0, \sigma | \mathbf{\hat{L}} |i,\sigma' \rangle \langle i, \sigma'| A \mathbf{\hat{S}} \cdot \mathbf{\hat{L}} | 0, \sigma \rangle}{E_0 - E_{i, \sigma'}} \notag \\
    \langle 0^{(1)}|  \hat{L}_p | 0^{(1)} \rangle \approx&2A \sum_{iq}  \frac{\langle 0 | \hat{L}_p |i \rangle \langle i| \hat{L}_q | 0 \rangle}{E_0 - E_{i}} \langle \sigma \sum_{\sigma'}|\sigma'\rangle \langle \sigma'|\hat{S}_q|\sigma\rangle \notag \\
    =&2A \sum_q \langle \sigma |\hat{S}_q|\sigma\rangle \sum_{i}  \frac{|\langle 0 | \hat{L}_p |i \rangle \langle i | \hat{L}_q |0 \rangle|}{E_0 - E_{i}}.
\end{align}
Therefore, $\langle\mathbf{\hat{L}}\rangle \approx c\langle\mathbf{\hat{S}}\rangle $ where
\begin{equation}
    c_{pq} = 2A \sum_{i}  \frac{|\langle 0 | \hat{L}_p |i \rangle \langle i | \hat{L}_q |0 \rangle|}{E_0 - E_{i}}.
\end{equation}
Comparing Eq.~\ref{eq:zeeman} and Eq.~\ref{eq:eff}, we find
\begin{equation} \label{eq:g&L}
    g_\mathrm{eff}^{pq} = g_e + 2A \sum_{i}  \frac{|\langle 0 | \hat{L}_p |i \rangle \langle i | \hat{L}_q |0 \rangle|}{E_0 - E_{i}}
\end{equation}
and thus the anisotropic part of the $g$-factor $g_\mathrm{eff} -g_e $ is basically $\langle\mathbf{\hat{L}}\rangle$ evaluated on the BO+SOC eigenstates. 

The spin-rotation coupling constant is related to this $\langle\mathbf{\hat{L}}\rangle$ by Eq.~\ref{Eq:E2_PT}. Combining this relationship with Eq.~\ref{eq:g&L} and the form of the model Hamiltonian $\hat{H} = \hat{\mathbf{S}} \cdot \mathbf{\epsilon} \cdot \hat{\mathbf{N}}$ yields Curl's 
relationship\cite{curl_jr_relationship_1965} between the spin-rotation coupling constant and the $g$-factor anisotropy,
$$\mathbf{\epsilon} =- \mathbf{I}^{-1}(g_e \mathcal{I}_3 - \mathbf{g})\hbar^2 $$

\section{From ``one-dimensional'' rotations to rotations in three dimensions}~\label{Sec:supp_3Dmodel}
For each phase space calculation performed in the main text, we investigated the effect of rotating around one axis in the body frame, usually one of the molecular principal axes, and treated that nuclear rotation classically. That being said, in reality, not only do molecules rotate within the body frame, but also the body frame (attached to the molecule) rotates within the lab frame. Moreover,  nuclear rotations are most accurately described quantum mechanically (not classically). In this section, we describe how we can combine multiple phase space calculations for rotations about different axes to describe freely rotating molecules in 3D quantum mechanically. Specifically, we will focus on symmetric top molecules and calculate the spin-rotation splitting between molecular eigenstates (electrons plus nuclei). 

We start by writing a model Hamiltonian for the molecular rotation and spin-rotation coupling whose parameters have been calculated from phase space calculations, each associated with one rotation axis in the body frame. The model Hamiltonian has the following form,
\begin{equation}\label{eq:supp_Hrot}
    H = \sum_{\mu=a,b,c} \frac{(\hat{\mathbf{N}}_\mu - \alpha_\mu \hat{\mathbf{S}}_\mu)^2}{2I_\mu}
\end{equation}
where $a,b$ and $c$ are the molecular principal axis, $c$ is the symmetry axis with the highest rotational symmetry,  $\mathbf{N}$ is the nuclear rotation angular momentum, $\mathbf{S}$ is the electronic spin angular momentum, $\alpha_\mu$ is a unit less constant factor, and $I$ is the inertia. Note that both $\hat{\mathbf{N}}$ and $\hat{\mathbf{S}}$ are quantum mechanical operators. The $\alpha_\mu$ is related to the classical $N_\mu=L^\mathrm{n,min}_{\mu}$ at the phase space minima for a rotation around the $\mu$ axis. For molecules with $S=1/2$, $\alpha_\mu$ is calculated  as $\alpha_\mu = 2L^\mathrm{n,min}_{\mu}/\hbar$. If we write the spin states of the two PS PESs at $N>0$ for rotation around $\mu$ as $|+\mu\rangle$ and $|-\mu\rangle$, we find that the Hamiltonian clearly reduces to the same form as was used recently for $\Lambda$-doubling~\cite{peng2026phasespaceelectronicstructure} in the nuclear rotational space without spin,
\begin{equation}
    H = \sum_{\mu=a,b,c} \frac{(\hat{\mathbf{N}}_\mu - L^\mathrm{n,min}_{\mu})^2}{2I_\mu}.
\end{equation}

To calculate the spin-rotation splitting, we need to diagonalize the Hamiltonian in Eq.~\ref{eq:supp_Hrot}. We will do this by writing out in the matrix form the Hamiltonian in the basis of $|N, M_N, S, M_s\rangle$, i.e. the eigenbasis of $\hat{\mathbf{N}}^2, \hat{N}_z, \hat{\mathbf{S}}^2$ and $\hat{S}_z$. Note that it is important to use the $|N,M_N, S, M_s\rangle$ basis, which is stationary in the lab frame, instead of the $|N,K_N, S, K_s\rangle$ basis (eigenbasis of $\hat{\mathbf{N}}^2, \hat{N}_c, \hat{\mathbf{S}}^2$ and $\hat{S}_c$) which rotates with the body frame, because we are looking for eigenstates that are stationary in the lab frame, with conserved $J$ and $J_z$ quantum numbers. When the Hamiltonian is anisotropic in the body frame, as in our case, we need to carefully treat how the Hamiltonian evolves when we change molecular orientation, a feat we can achieve by working in the lab frame. (If we were to diagonalize the Hamiltonian in the body frame basis, we would only capture the rotation of the molecule relative to the body frame and ignore the rotation of the orientation for a freely-rotating molecule in the lab frame.) 

The key term in the Hamiltonian that captures the spin-rotation splitting is the cross term between $\hat{\mathbf{N}}$ and $\hat{\mathbf{S}}$,
\begin{equation}
\hat{H}^\mathrm{SR} = -\sum_{\mu=a,b,c} \frac{\alpha_\mu}{I_\mu} \hat{\mathbf{N}}_\mu \hat{\mathbf{S}}_\mu = \hat{\mathbf{N}}\cdot \epsilon \cdot \hat{\mathbf{S}}
\end{equation}
where the spin-rotation coupling constant $\epsilon$ is a diagonal $3\times3$ matrix with the diagonal elements, and we can identify
\begin{eqnarray} \label{eq:alpha_epsilon}
    \epsilon_\mu=-\alpha_\mu / I_\mu.
\end{eqnarray} 
For an oblate symmetric top molecule such as CH$_3$, $\alpha_a = \alpha_b$, $I_a = I_b$, and $\epsilon_a=\epsilon_b$ due to molecular symmetry. 
Therefore, we can separate $\hat{H}^\mathrm{SR}$ into the isotropic part and the anisotropic part,
\begin{equation}
    \hat{H}^\mathrm{SR} = \epsilon_b \hat{\mathbf{N}}\cdot \hat{\mathbf{S}} - (\epsilon_b - \epsilon_c) \hat{\mathbf{N}}_c \hat{\mathbf{S}}_c.
\end{equation}
Consider now a given ($N$, $K$) rotational level, where $K$ is the projection of $N$ on the body frame $c$ axis. 
Assuming $\mathbf{S}$ is strongly coupled to $\mathbf{N}$ and both point in the same direction in a classical sense, as should be the case for small molecules including CH$_3$, one can effectively reduce the $c$-axis component to 
\begin{equation}
    \hat{\mathbf{N}}_c \hat{\mathbf{S}}_c = \frac{K^2}{N(N+1)} \hat{\mathbf{N}}\cdot \hat{\mathbf{S}}
\end{equation}
for the given ($N, K$) rotational level; see
Refs.~\cite{sutter2006molecular, gordy1984microwave,flygare1974magnetic} for more details and a proper quantum treatment in terms of the so-called direction cosine matrix approach. Thus, we can rewrite the $\hat{H}^\mathrm{SR}$ as an effective isotropic Hamiltonian   
\begin{equation} \label{eq:sr_model}
    \hat{H}^\mathrm{SR} = \left[\epsilon_b - (\epsilon_b - \epsilon_c) \frac{K^2}{N(N+1)} \right]\hat{\mathbf{N}}\cdot \hat{\mathbf{S}}.
\end{equation}
Eq.~\ref{eq:sr_model} is very helpful because $\hat{\mathbf{N}}\cdot \hat{\mathbf{S}}$ can be easily diagonalized within a lab frame basis. Note that we have not made any assumption that the Hamiltonian is isotropic, which is clear because the effective isotropic coupling constant 
\begin{equation}
    \gamma_\mathrm{eff} = \epsilon_b - (\epsilon_b - \epsilon_c) \frac{K^2}{N(N+1)}
\end{equation}
depends on $K$; in other words,  the Hamiltonian is still anisotropic.

Finally, to finish the calculation, we must diagonalize  $\hat{\mathbf{N}} \cdot \hat{\mathbf{S}}$ (which is an isotropic operator). 
One can easily work out the energy splitting expression from the above Hamiltonian using operator algebra and noting that
\begin{equation}
    \hat{\mathbf{N}} \cdot \hat{\mathbf{S}} = 1/2\left[(\hat{\mathbf{N}} + \hat{\mathbf{S}})^2 - \hat{\mathbf{N}}^2 - \hat{\mathbf{S}}^2 \right].
\end{equation}
Clearly, the energy difference between states with $J=N+1/2$ and $J=N-1/2$ yields the spin-rotation splitting. In particular, for $N=1$ and $K=1$, the $6\times6$ matrix yields one eigenvalue with fourfold degeneracy ($J=3/2$) and one eigenvalue with twofold degeneracy ($J=1/2$). The difference between these two eigenvalues is the spin-rotation splitting. 
In particular, note that the quantum mechanical energy expectation value for the $J=N+1/2$ state is
\begin{align}
    \langle \hat{H}^\mathrm{SR}\rangle_{N+1/2} &= 1/2 \left[\epsilon_b - (\epsilon_b - \epsilon_c) \frac{K^2}{N(N+1)} \right]\left[(N+\frac{1}{2})(N+\frac{3}{2}) - N(N+1) - \frac{1}{2} \cdot \frac{3}{2} \right]\\
    &= 1/2 \left[\epsilon_b - (\epsilon_b - \epsilon_c) \frac{K^2}{N(N+1)} \right] N,
\end{align}
and for the $J=N-1/2$ state
\begin{align}
    \langle \hat{H}^\mathrm{SR}\rangle_{N-1/2} &= 1/2 \left[\epsilon_b - (\epsilon_b - \epsilon_c) \frac{K^2}{N(N+1)} \right]\left[(N-\frac{1}{2})(N+\frac{1}{2}) - N(N+1) - \frac{1}{2} \cdot \frac{3}{2} \right]\\
    &= 1/2 \left[\epsilon_b - (\epsilon_b - \epsilon_c) \frac{K^2}{N(N+1)} \right] (-N-1).
\end{align}
Therefore, the energy difference between these two states is 
\begin{equation}
     \Delta E^\mathrm{SR} = \langle \hat{H}^\mathrm{SR}\rangle_{N+1/2}  -  \langle \hat{H}^\mathrm{SR}\rangle_{N-1/2} =(N+1/2)\left[ \epsilon_b - (\epsilon_b - \epsilon_c) K^2/N(N+1)\right].
\end{equation}
which matches the analytical expression that was used to fit the spin-rotation coupling constant for the experimental spectra in  Ref.~\citenum{davis1997jet} (see Eq. 4  therein). 

\section{Meaning of phase space nuclear momentum} \label{sec:supp_P}
In the main text, we pointed out that the expectation value of the nuclear momentum operator $\langle \hat{\mathbf{P}}\rangle$ in the Born-Huang framework should be understood as the total nuclear and electronic momentum; the same assignment should hold for the nuclear momentum parameter $\mathbf{P}$ in within a phase space electronic structure framework as well. In this section, we review the mathematical argument behind this interpretation from Ref.~\citenum{littlejohn2024diagonalizing}, and provide an intuitive understanding. 

Mathematically, the key to understanding the meaning of $\langle \hat{\mathbf{P}}\rangle$ is to recognize that the operator $\hat{\mathbf{P}}$ in the Born-Huang framework acts only in the nuclear space and not on the electronic adiabatic states. We know that applying the nuclear momentum operator to the total wavefunction yields the nuclear momentum, while applying the total nuclear and electronic momentum operators to the total wavefunction yields the total nuclear and electronic momentum. In the Born-Huang framework, we expand the total nuclear and electronic wavefunction $\Psi(\mathbf{X}, \mathbf{\mathbf{r}}) = \sum_n \psi_n (\mathbf{X}) \phi_n(\mathbf{X}, \mathbf{\mathbf{r}})$ as products of nuclear wavefunctions $ \psi_n (\mathbf{X})$ and electronic wavefunctions written in the frame of the nuclei $\phi_n(\mathbf{X}, \mathbf{\mathbf{r}})$ that depends only on the distance between $\mathbf{X}$ and $\mathbf{r}$ but not each of $\mathbf{X}$ and $\mathbf{r}$  separately. In this case, one can show that applying the total nuclear and electronic momentum operators to the total wavefunction 
\begin{align}
    (\hat{\mathbf{P}} + \hat{\mathbf{p}})\Psi(\mathbf{X}, \mathbf{\mathbf{r}}) = & \sum_n \left[(\hat{\mathbf{P}} + \hat{\mathbf{p}}) \psi_n (\mathbf{X})\right]\phi_n(\mathbf{X}, \mathbf{r})  + \psi_n (\mathbf{X}) \left[(\hat{\mathbf{P}} + \hat{\mathbf{p}})\phi_n(\mathbf{X}, \mathbf{r}) \right] \notag \\
    =& \sum_n \left(\hat{\mathbf{P}} \psi_n (\mathbf{X})\right)\phi_n(\mathbf{X}, \mathbf{r}) \label{eq:supp_apply_PP}
\end{align}
is in fact equivalent to applying the nuclear momentum operator to the nuclear-only wavefunction. Therefore,  $\langle\hat{\mathbf{P}}\rangle$ in the Born-Huang framework represents the total nuclear and electronic momentum.  In  Eq.~\ref{eq:supp_apply_PP} above, we have used the condition 
\begin{equation} 
    \hat{\mathbf{P}} \phi_n = -\hat{\mathbf{p}}\phi_n = i\hbar \sum_i \nabla_i\phi_n. 
\end{equation}

The above mathematical argument can also be understood intuitively. The key ingredient in the proof above is that the electronic wavefunction is only a function of $\mathbf{r} - \mathbf{X}$ but not $\mathbf{r} $ or $ \mathbf{X}$ independently. Intuitively, this parametrization means electrons are attached to the nuclei, so when we move the nuclei while fixing the electronic wavefunction, the electrons are actually moving with the nuclei in space. Therefore, the $\langle\hat{\mathbf{P}}\rangle$ carries not only the momentum of nuclei, but also the momentum of electrons.

\section{Definition and symmetry of phase space $\hat{\mathbf{\Gamma}}$ operator} \label{sec:supp_gamma}
As discussed in the main text, the phase space $\hat{\mathbf{\Gamma}}$ operator contains up to two terms, $\hat{\mathbf{\Gamma}}_A = \hat{\mathbf{\Gamma}}_A^\mathrm{spatial} + \hat{\mathbf{\Gamma}}_A^\mathrm{spin}$. The first spatial term can be further divided into two terms,  $\hat{\mathbf{\Gamma}}_A^\mathrm{spatial} = \hat{\mathbf{\Gamma}}_A' + \hat{\mathbf{\Gamma}}_A''$. 
\begin{itemize}
\item $(i)$ $\hat{\mathbf{\Gamma}}_A'$ is the electron translation factor 
\begin{gather}
    \hat{\mathbf{\Gamma}}'_A = \frac{1}{2i\hbar}(\hat{\Theta}_A \hat{\mathbf{p}} +\hat{\mathbf{p}} \hat{\Theta}_A),
\end{gather}
where the electronic linear momentum is partitioned to each nucleus according to a spatial locality factor as a function of the distance between the electron and the nucleus
\begin{eqnarray}  
\label{eq:theta}
    \hat{\Theta}_A(\mathbf{x}) = \frac{Q_A e^{- |\hat{\mathbf{x}} - \mathbf{X}_A|^2/\sigma^2}}{\sum_B Q_B e^{- |\hat{\mathbf{x}} - \mathbf{X}_B|^2/\sigma^2}}.
\end{eqnarray}
Here, $\sigma$ is a parameter that captures the length over which atomic density changes. In theory, the prediction of spin-rotation coupling strength when nuclei are only rotating and not vibrating should be largely independent of this $\sigma$ parameter~\cite{peng2026phasespaceelectronicstructure}. In practice, all PS calculations in this paper use $\sigma=3.27$. 

\item $(ii)$  $ \hat{\mathbf{\Gamma}}''_A$ is the electron orbital rotation factor
\begin{gather}
    \hat{\mathbf{\Gamma}}''_A = \sum_B \zeta_{AB}(\mathbf{X}_A - \mathbf{X}_B^0) \times (\mathbf{K}_B^{-1} \hat{\mathbf{J}}^{(l)}_B), \\
    \hat{\mathbf{J}}_B^{(l)} = \frac{1}{2i\hbar}\left( (\hat{\mathbf{x}} - \mathbf{X}_B) \times (\hat{\Theta}_B \hat{\mathbf{p}})  + (\hat{\mathbf{x}} - \mathbf{X}_B) \times (\hat{\mathbf{p}}\hat{\Theta}_B ) \right),
\end{gather}
where $ \mathbf{X}_B^0 = \frac{\sum_A\zeta_{AB} \mathbf{X}_A}{\sum_A \zeta_{AB}}$ is a local average position, 
\begin{equation}
    \mathbf{K}_B = \sum_A \zeta_{AB} \left((\mathbf{X}_A^\top \mathbf{X}_A - \mathbf{X}_B^{0\top} \mathbf{X}_B^0) \mathcal{I}_3 - (\mathbf{X}_A \mathbf{X}_A^\top - \mathbf{X}_B^0 \mathbf{X}_B^{0\top}) \right)
\end{equation}
is a local version of moment of inertia partitioned to each atom, $\mathcal{I}_3$ is the $3\times3$ identity matrix, and $\zeta_{AB}=M_A e^{-|\mathbf{X}_A - \mathbf{X}_B|^2/8\sigma^2}$ is a locality factor that reflects the localization of electrons across different atoms. 

\item $(iii)$ $\hat{\mathbf{\Gamma}}_A^\mathrm{spin}$, also called $\hat{\mathbf{\Gamma}}'''_A$ or $\hat{\mathbf{\Gamma}}''^{(s)}$ in previous phase space literature such as Refs.~\citenum{bradbury2025symmetry,peng2026phasespaceelectronicstructure,bian2025review}, is the electron spin rotation factor that is defined similarly to the electron orbital rotation factor $\hat{\mathbf{\Gamma}}''_A$ except for the electronic spin 
\begin{gather}
    \hat{\mathbf{\Gamma}}_A^\mathrm{spin} = \sum_B \zeta_{AB}(\mathbf{X}_A - \mathbf{X}_B^0) \times (\mathbf{K}_B^{-1} \hat{\mathbf{J}}^{(s)}_B), \\
    \hat{\mathbf{J}}_B^{(s)} = \frac{1}{i\hbar} \hat{\mathbf{S}} \hat{\Theta}_B.
\end{gather}
\end{itemize}

The $\hat{\mathbf{\Gamma}}$ operator defined above satisfies some important symmetries of the nonadiabatic derivative coupling. When only the spatial part of electrons is described in the molecular frame (and the spin part is described in the space frame), $\hat{\mathbf{\Gamma}}_A \equiv \hat{\mathbf{\Gamma}}_A^\mathrm{spatial}$  satisfies the following constraints: 
\begin{gather}
    -i\hbar \sum_A \hat{\mathbf{\Gamma}}_A + \hat{\mathbf{p}} = 0, \label{supp_sym1}\\
    \left[ -i\hbar{\sum_B \frac{\partial}{\partial \mathbf{X}_B}} + \mathbf{\hat{p}}, \hat{\mathbf{\Gamma}}_A\right] = 0, \label{supp_sym2}\\
    -i\hbar \sum_A \mathbf{X}_A \times \hat{\mathbf{\Gamma}}_A + \hat{\mathbf{L}}= 0, \label{supp_sym3}\\
    \left[ -i\hbar \sum_B \left(\mathbf{X}_B \times \frac{\partial}{\partial \mathbf{X}_B}\right)_\beta + \hat{L}_\beta + \hat{S}_\beta, \hat{\Gamma}_{A\gamma} \right] = i\hbar \sum_\alpha \epsilon_{\alpha \beta \gamma} \hat{\Gamma}_{A\alpha}. \label{supp_sym4}
\end{gather} 
When both the spatial part and the spin part of electrons are described in the molecular frame, $\hat{\mathbf{\Gamma}}_A \equiv \hat{\mathbf{\Gamma}}_A^\mathrm{spatial}+\hat{\mathbf{\Gamma}}_A^\mathrm{spin} $, and Eq.~\ref{supp_sym3} becomes
\begin{equation} \label{supp_sym3_L+S}
    -i\hbar \sum_A \mathbf{X}_A \times \hat{\mathbf{\Gamma}}_A + \hat{\mathbf{L}} + \hat{\mathbf{S}} = 0.
\end{equation}
Eq.~\ref{supp_sym1} and Eq.~\ref{supp_sym3} (or Eq.~\ref{supp_sym3_L+S}) enforce the translational and rotational invariance of the total system, and Eq.~\ref{supp_sym2} and Eq.~\ref{supp_sym4} ensure $\hat{\mathbf{\Gamma}}_A$ operator is translationally and rotationally invariant.

\section{Alternative phase space approaches to estimate spin-rotation splitting}\label{sec:supp_alternative_dE}

In the main body of the text above, we have calculated the spin-rotation splitting for a given rotational axis (say the $\mu$ direction) by measuring the vertical energy gap at $L_\mu=\hbar$ in Fig.~\ref{fig:PES}. 
As an alternative means to extract a spin-rotation coupling constant, following ``Method I'' or ``Method II'' from Ref.~\citenum{peng2026phasespaceelectronicstructure}, one can model the two phase space PESs in Fig.~\ref{fig:PES} as two parabolas $E_{\mu\mu}(L^\mathrm{n}_\mu)=\frac{1}{2I_\mu}(L^\mathrm{n}_\mu\pm\alpha S_\mu)^2$, so that  
$\alpha$ can be ascertained by fitting the location of the minimum of the PES with the $+\mu$ spin character, $L^\mathrm{n,min}_\mu$. 
Furthermore, based on the stationary condition of a PS PES, this minimum point satisfies $L^\mathrm{n,min}_\mu = \langle \hat{L}_{\mu}\rangle|_{L^\mathrm{n,min}_\mu}$,
where the bracket $\langle ~\rangle$ now indicates expectation value with respect to the ground-state surface. For an $S=1/2$ system, $\alpha_\mu = 2L^\mathrm{n,min}_{\mu}/\hbar$ (and then the spin-rotation coupling constant $\epsilon_{\mu\mu}$ can again be calculated as before with Eq. \ref{eq:alpha_epsilon}).
 In practice, when $\langle \hat{L}_{\mu}\rangle$ is small (because $\langle \hat{L}_\mu\rangle$ is quenched under the nonrelativistic BO approximation for most polyatomic molecules), we can further approximate the location of the minima using $\langle \hat{L}_{\mu}\rangle$ evaluated on the relevant eigenstate that is produced at $\mathbf{L}^\mathrm{n} = 0$ (i.e. without any external nuclear rotation) and with the corresponding spin character,  i.e. $\alpha_{\mu} \approx 2\langle \hat{L}_{\mu}\rangle|_{L^\mathrm{n}_\mu\rightarrow0^+}/\hbar$. The above discussion is framed for symmetric top molecules but can be easily generalized to asymmetric top molecules.  Empirically, if we were to follow the approaches in this paragraph, one finds very similar results as found in the main body of the text.

\section{Phase space calculation with $\hat{\mathbf{\Gamma}}^\mathrm{spin}$} \label{sec:gamma3}

\begin{figure}[htb!] 
    \centering
    \includegraphics[width=0.5\columnwidth]{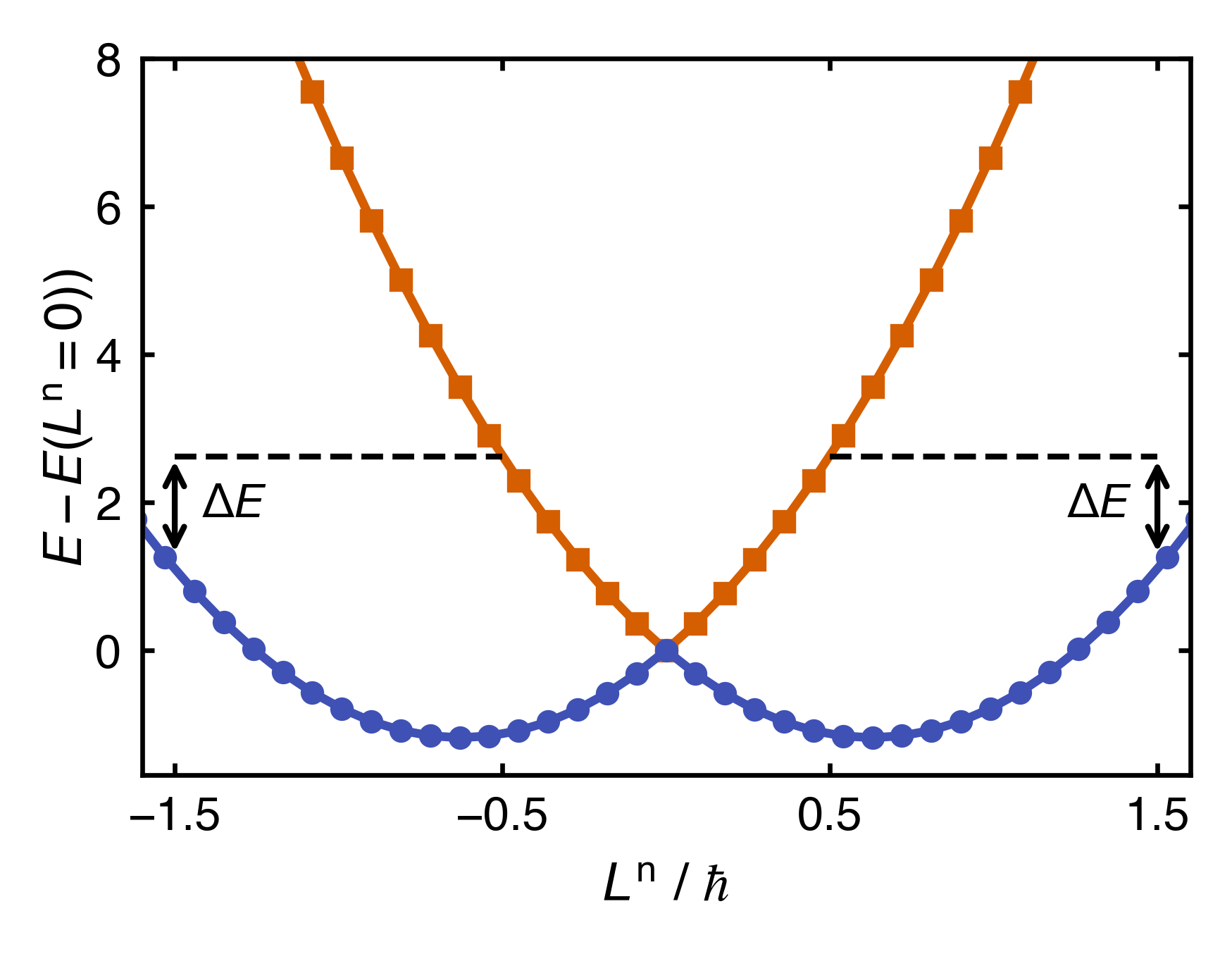}
    \caption{Lowest two potential energy surfaces from the phase space theory, plotted as a function of the canonical nuclear momentum $L^\mathrm{n}$ and colored based on their relationship to the total electronic and nuclear eigenstates. Figure not to scale. Intuitively, the spin-rotation splittings for $N=1$ arise from the difference in energy between the broken symmetry ground state surface (with spin aligned with the rotation) at $L^\mathrm{n}=\pm 1.5$ and the excited state surface (with spin antialigned with the rotation) at $L^\mathrm{n}=\pm0.5$, where the energy difference is labeled as $\Delta E$.}
    \label{fig:supp_PES_G3}
\end{figure}

As discussed in the main body of the text, when running phase space electronic structure calculations, there is always the question of whether or not to put the electronic spin in the molecular frame or in the lab frame. While in the main manuscript, we chose the former, if one chooses the latter by including $\hat{\mathbf{\Gamma}}^\mathrm{spin}$, one can still extract spin rotation couplings but with a caveat: One must now compare different energies at different values of $L^\mathrm{n}$, i.e. $L^\mathrm{n}=N-1/2$ and $L^\mathrm{n}=N+1/2$. See Fig.~\ref{fig:supp_PES_G3}.

Physically, whether or not to include $\hat{\mathbf{\Gamma}}^\mathrm{spin}$ must depend on how strong the SOC is and how slow the nuclear rotations are. For weak SOC and fast nuclear rotations, one must expect that including $\hat{\mathbf{\Gamma}}^\mathrm{spin}$ will lead to large errors relative to experiment. Vice versa, for larger SOC and fast nuclear rotations,
including $\hat{\mathbf{\Gamma}}^\mathrm{spin}$ should lead to smaller errors (as, in the extreme limit of infinite SOC and infinitesimally slow nuclear rotations, the spin should entirely align with the body frame). This intuition is confirmed below in Fig.~\ref{fig:supp_G3}.

\begin{figure}[htb!] 
    \centering
    \includegraphics[width=0.5\columnwidth]{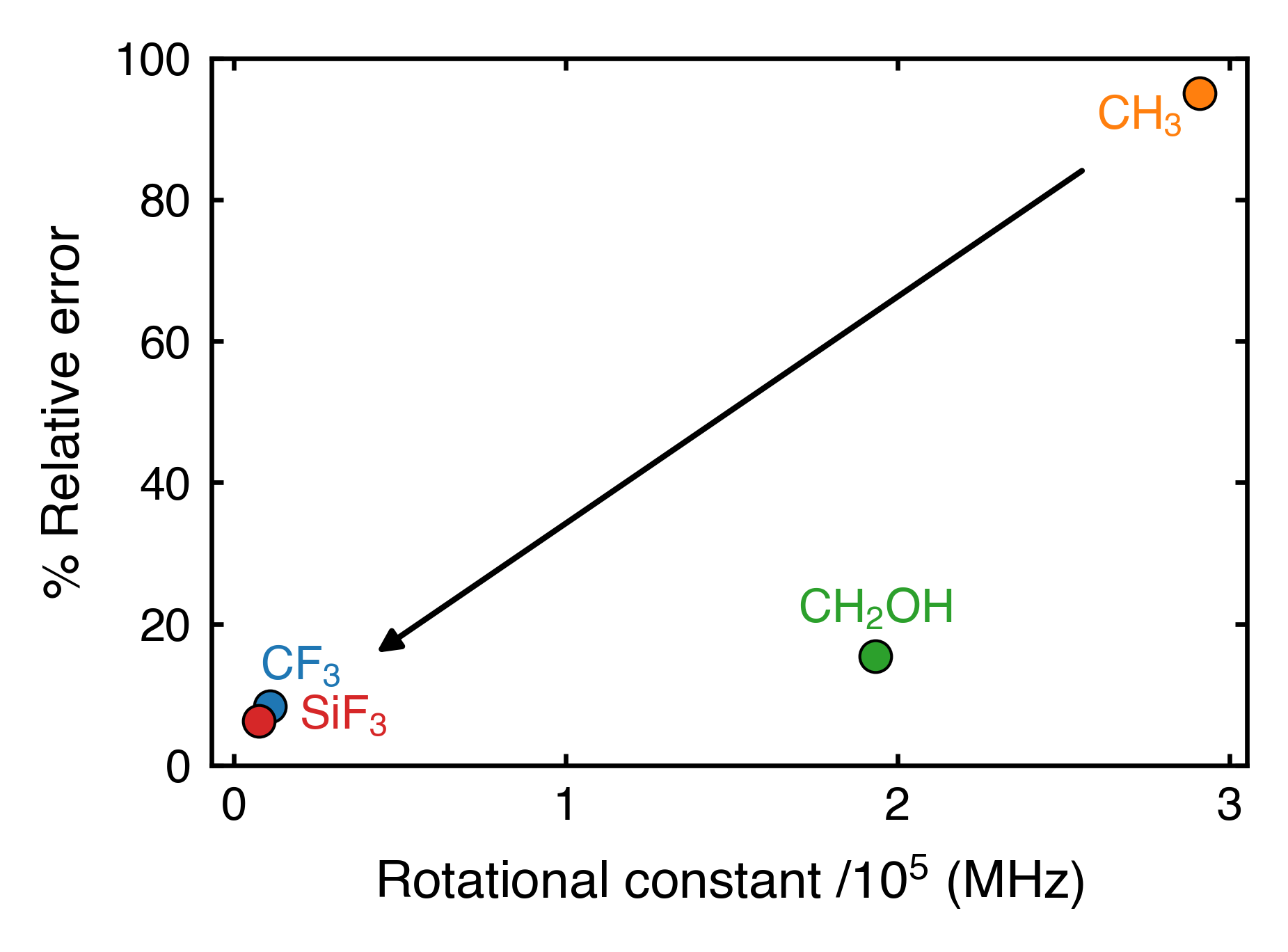}
    \caption{The error of major spin-rotation coupling constants of each molecule predicted using a phase space Hamiltonian that includes the $\hat{\mathbf{\Gamma}}^\mathrm{spin}_A (\mathbf{X})$ term, relative to the values predicted without $\hat{\mathbf{\Gamma}}^\mathrm{spin}_A (\mathbf{X})$. The relative errors become smaller for heavier molecules with smaller rotational constants.}
    \label{fig:supp_G3}
\end{figure}

\section{Full spin-rototation tensor of CH$_2$OH} \label{sec:supp_ch2oh}
The full $3\times 3$ spin-rotation tensor of CH$_2$OH in the principal axis basis is listed below:
\[
\boldsymbol{\epsilon}~ \mathrm{ (MHz)}
=
\begin{pmatrix}
-467.5 & -38.1 & 5.0 \\
-114.7 & -117.9 & 0.7 \\
-39.6 & -3.3 & 0.2
\end{pmatrix}.
\]
The conventional spectroscopic $\epsilon$ only contains the symmetric part of the spin-rotation tensor. Therefore, we also report the symmetrized tensor below:
\[
\boldsymbol{\epsilon}_\mathrm{sym}~ \mathrm{ (MHz)}=
\begin{pmatrix}
-467.5 & -76.4 & -17.3 \\
-76.4 & -117.9 & 1.3 \\
-17.3 & -1.3 & 0.2
\end{pmatrix}.
\]


\putbib
\end{bibunit}

\end{document}